\makeatletter
\@ifundefined{@parse@version@dash}{%
\def\@parse@version#1{\@parse@version@0#1}
\def\@parse@version@#1/#2/#3#4#5\@nil{%
\@parse@version@dash#1-#2-#3#4\@nil}
\def\@parse@version@dash#1-#2-#3#4#5\@nil{%
  \if\relax#2\relax\else#1\fi#2#3#4 }
}{}
\makeatother

\documentclass[aps,prx,twocolumn,superscriptaddress]{revtex4-2}
\usepackage{epsfig,amsmath,amssymb,amsfonts,physics,graphicx,color,calc,epstopdf}
\usepackage[ruled,vlined]{algorithm2e}

\usepackage{xcolor,hyperref}
\hypersetup{
   colorlinks,
   linkcolor={blue!50!black},
   citecolor={blue!50!black},
   urlcolor={blue!80!black}
}

 \let\b=\beta

\def\ie{{\em i.e. }}
\def\eg{{\em e.g. }}
\def\cf{{\em cf. }}

 \def\JJ{{\cal J}} 
 
\def\KK{{\cal K}}

\def\de{\mathrm d}

\def\to{\rightarrow}

\newcommand{\beq}{\begin{equation}} \newcommand{\eeq}{\end{equation}}

\begin{document}

\title{Sparse generative modeling
via parameter-reduction of Boltzmann machines:\\
application to protein-sequence families}

\author{Pierre Barrat-Charlaix} 
\affiliation{Biozentrum, Universit\"at Basel, Switzerland - Swiss Institute of Bioinformatics, 4056 Basel, Switzerland}

\author{Anna Paola Muntoni}
\affiliation{Department of Applied Science and Technology (DISAT), Politecnico di Torino, Corso Duca degli Abruzzi 24, I-10129 Torino, Italy}
\affiliation{Italian Institute for Genomic Medicine, IRCCS Candiolo, SP-142, I-10060 Candiolo (TO), Italy}
\affiliation{Sorbonne Universit\'e, CNRS, Institut de Biologie Paris Seine, Biologie Computationnelle et Quantitative – LCQB, 75005 Paris, France}
\affiliation{Laboratoire de Physique de l'Ecole Normale Sup\'erieure, ENS, Universit\'e PSL, CNRS, Sorbonne Universit\'e, Universit\'e de Paris, F-75005 Paris, France}

\author{Kai Shimagaki} 
\affiliation{Sorbonne Universit\'e, CNRS, Institut de Biologie Paris Seine, Biologie Computationnelle et Quantitative – LCQB, 75005 Paris, France}

\author{Martin Weigt}
\affiliation{Sorbonne Universit\'e, CNRS, Institut de Biologie Paris Seine, Biologie Computationnelle et Quantitative – LCQB, 75005 Paris, France}

\author{Francesco Zamponi}
\affiliation{Laboratoire de Physique de l'Ecole Normale Sup\'erieure, ENS, Universit\'e PSL, CNRS, Sorbonne Universit\'e, Universit\'e de Paris, F-75005 Paris, France}

\begin{abstract} 
Boltzmann machines (BM) are widely used as generative models.
For example, pairwise Potts models (PM), which are instances of the BM class, provide accurate statistical models of families of evolutionarily related protein sequences. Their parameters are the local fields, which describe site-specific patterns of amino-acid conservation, and the two-site couplings, which mirror the coevolution between pairs of sites. This coevolution reflects structural and functional constraints acting on protein sequences during evolution. The most conservative choice to describe the coevolution signal is to include all possible two-site couplings into the PM. This choice, typical of what is known as Direct Coupling Analysis, has been successful for predicting residue contacts in the three-dimensional structure, mutational effects, and in generating new functional sequences.
However, the resulting PM suffers from important over-fitting effects: many couplings are small, noisy and hardly interpretable; the PM is close to a critical point, meaning that it is highly sensitive to small parameter perturbations.
In this work, we introduce a general parameter-reduction procedure for BMs, 
via a controlled iterative decimation of the less statistically significant couplings, identified by an information-based criterion that selects either weak or statistically unsupported couplings. For several protein families, our procedure allows one to remove more than $90\%$ of the PM couplings, while preserving the predictive and generative properties of the original dense PM,
and the resulting model is far away from criticality, hence more robust to noise.
\end{abstract}

\maketitle 

\section{Introduction}

Many applications of generative modeling, especially in biological systems, are confronted to a limited amount of available data, from which a large number of parameters have to be inferred~\cite{nguyen2017inverse}. 
A particularly interesting example is that of
proteins, which belong to the most interesting complex systems in nature and are essential in almost all biological processes. Most of them robustly fold into well-defined three-dimensional structures, which in turn form the basis of their functionality. This triangular sequence-structure-function relationship has, over several decades now, attracted substantial attention in biological physics~\cite{bryngelson1995funnels,dill2012the}.

A fascinating approach to the generative modeling of biological sequences has emerged over the last years \cite{de2013emerging,cocco2018inverse}. In the course of evolution, biological sequences accumulate mutations and become more diverse. We can now easily observe the sequence variability across large families of so-called homologous proteins, {\em i.e.}~proteins of common evolutionary ancestry and of close to equivalent function but in different species or biological pathways~\cite{el2019pfam}. Such homologous proteins may differ by 70-80\% of their amino acids without substantial changes in structure and function. However, their sequence variability is not fully random: a vast majority of mutations is deleterious, reducing protein stability or functionality. They are thus suppressed by natural selection. Only protein variants of similar or even better functionality are maintained.
In this way, the protein's structure and function constrain the viable sequence space that can be explored by evolution. Inverting this argument, the empirically observed variability of homologous sequences contains information about such evolutionary constraints, albeit frequently well hidden. 
This idea is at the basis of the concept of data-driven ``sequence landscapes'', {\em i.e.}~classes of models that describe the statistical properties of protein families, assigning high probabilities to functional amino-acid sequences and low probabilities to non-functional ones \cite{levy2017potts,cocco2018inverse}. The log-probability (or minus ``energy'') is thus interpreted as a measure of sequence
fitness, hence the name of sequence (fitness) landscape \cite{de2014empirical}.
Among the best known such models are Potts models (PM), parameterized by local fields and two-site interaction couplings (\cf~below for details), and 
constructed via the Direct Couplings Analysis (DCA) method, which is now firmly established~\cite{levy2017potts,cocco2018inverse}. 
The DCA parameters can be obtained via inference or learning procedures ~\cite{balakrishnan2011learning,ekeberg2014fast, ekeberg2013improved, figliuzzi2018pairwise}, and they can be used to extract useful information on molecular structure~\cite{weigt2009identification, morcos2011direct,marks2012protein,ovchinnikov2017protein} and function ~\cite{morcos2014coevolutionary,szurmant2018inter}, on the effects of mutations ~\cite{figliuzzi2015coevolutionary,hopf2017mutation}, and to generate new artificially-designed molecules with specific properties~\cite{tian2018co,russ2020an}.

A concrete implementation of DCA is the following~\cite{figliuzzi2018pairwise}.
Given training data in the form of a Multiple Sequence Alignment (MSA) of $M$ homologous sequences of aligned length $L$, the PM parameters are learned by the so-called Boltzmann machine learning (BML) algorithm \cite{ackley1985learning}. By performing gradient ascent on the log-likelihood of the model given the data, BML determines values of couplings and fields such that the one- and two-site model frequencies match the empirical ones derived from the MSA. A standard pairwise $q$-state PM is thus specified by $q^2 L(L-1)/2$ couplings and $q L$ fields, where, for proteins, $q = 21$ corresponds to the 20 naturally occurring amino acids plus the gap symbol used for insertions or deletions.

Crucially, despite the fact that
modern sequencing techniques are making available a huge amount of biological sequences, and in particular hundreds of millions of protein sequences \cite{uniprot2019}, a serious over-fitting problem is present when PM are used as models of protein families. 
In fact, with typical sequence lengths $L\sim 50-500$, the parameters to be inferred are $\sim 10^6-10^8$, which in most cases substantially exceeds 
the available information from the MSA.
The resulting over-fitting is manifested in several ways: (i)~many couplings turn out to be rather small and noisy, (ii)~the PM is close to a critical point, {\em i.e.}~it can be very susceptible to small changes in its parameters, and (iii)~different training procedures, {\em e.g.}~with different initial conditions, can lead to significant changes in the sets of parameters without affecting the fitting accuracy, which severely limits the interpretability of the model.

These observations call for a parameter reduction procedure, which aims at identifying a minimal set of couplings needed to accurately describe the training data without overfitting.
Hopfield-Potts models~\cite{shimagaki2019selection} and the more general Restricted Boltzmann Machines (RBM)~\cite{tubiana2019learning} lead to a dimensional reduction of parameter space by learning collective ``patterns'' from sequence data, which in turn can be interpreted as extended sequence motifs and are activated via a limited number of hidden variables. The resulting coupling matrix is low-rank but still dense. A complementary approach aims at sparsifying the network of couplings: $\ell_1$-norm regularization has been used in a number of approximate methods~\cite{jones2012psicov,kamisetty2013assessing}, but cannot be easily used for generative modeling, because the regularization penalizes also non-zero couplings, which in turn assume too small values. Alternatively, a ``color-compression'' scheme~\cite{rizzato2020inference} has been proposed, which groups together sequence symbols with low frequency in specific sites. However, frequent symbols may also be involved into statistically non-supported couplings.  Another example is that proposed in \cite{gao2018correlation} where a candidate sparse graph topology is sought by pruning the MSA columns associated with low values of the mutual information. Although this method has to be preferred when $L$ is so large to prevent the standard DCA implementations, it completely loses some information on the target statistical model.
Overall, a statistically principled and efficient approach to construct sparse PM for protein sequence modeling is still lacking.

In this work, we introduce an information-theory based ``decimation'' procedure, which allows for an iterative and controlled removal of irrelevant couplings. 
As a result, parameters are removed either if they have no statistical support (as in color compression), or if they have statistical support for being very small.
We show that up to about 90\% of the coupling parameters can be removed without observing any substantial change in the fitting accuracy and in the generative properties of the resulting Sparse Potts Model (SPM).
Although greedy, our pruning scheme does not require to add extra terms in the energy function of the model, at variance with any treatable regularization, like  $\ell_1$ or $\ell_2$, and therefore it preserves the generative properties of PM.
Finally, we show that the resulting SPM are not close to criticality, at variance with the original PM learned using standard DCA. Our results thus demonstrate that the observed criticality of PMs inferred from protein sequence data is not an intrinsic feature of the biological systems themselves, \cf~\cite{mora2011biological}, but results from the over-fitting in the learning procedure.

\section{An information-guided decimation procedure} 

With each sequence $S = (s_1,\cdots,s_L)$ of length $L$, in which $s_i$ can take $q$ possible values ($q=21$ for proteins), a PM associates a statistical ``energy'' or Hamiltonian $H(S)$, written as a sum over single-site fields $h_i(s_i)$ and two-site couplings $J_{ij}(s_i,s_j)$:
\beq\label{eq:H}
H(S) = -\sum_{1\leq i<j \leq L} J_{ij}(s_i,s_j) - \sum_{1\leq i \leq L} h_i(s_i) \ .
\eeq
The negative of the Hamiltonian can be interpreted as a ``fitness score'' for protein sequence $S$, with an associated Boltzmann probability
$P(S) = \exp\{-H(S)\}/Z$, where $Z=\sum_S \exp\{-H(S)\}$ is the partition function guaranteeing correct normalization of $P$. Hence, the surface defined by $H(S)$ over the space of sequences can be interpreted as a ``fitness landscape'' or -- using a more cautious term -- ``sequence landscape'' for the protein family represented by the training MSA.  
We define the ``model density'' $d$ as the number of non-zero couplings  $J_{ij}(a,b)\neq 0$ divided by the total number of possible couplings $q^2 L(L-1)/2$. Note that this definition is given element-wise, {\em i.e.}~for each $i,j,a,b$, and not block-wise for entire $q\times q$ matrices $J_{ij}$ coupling two sites $i,j$.
Fields are not decimated and do not contribute to the model density: we consider them an essential ingredient of the model because they encode amino-acid conservation. 

A fully connected model, {\em i.e.}~with $d=100\%$, can be trained to arbitrarily high accuracy using standard BML~\cite{figliuzzi2018pairwise}. 
Let us define the empirical one-site frequency $f_i(a)$ of observing amino acid $a$ in position $i$ in the MSA, and two-site frequency $f_{ij}(a,b)$ of observing amino acid
$a$ in position $i$ and $b$ in position $j$ in the same sequence of the MSA. BML performs a gradient ascent on the log-likelihood, which gives update equations for the couplings and fields at each learning epoch: 
\beq\label{eq:PMlearning}
\begin{split}
\delta h_i(a) &= \eta_h [ f_i(a) - p_i(a) ] \ , \\
\delta J_{ij}(a,b) &= \eta_J [ f_{ij}(a,b) - p_{ij}(a,b) ] \ ,
\end{split}\eeq
where the $p_i(a), p_{ij}(a,b)$ are the one- and two-site marginal probabilities of the PM, which are estimated at each iteration of the learning by sampling $P(S)$ via a standard Markov Chain Monte Carlo (MCMC) simulation, and $\eta_h, \eta_J$ are the learning rates for fields and couplings. 
These equations are iterated until convergence to a fixed point, at which the model almost perfectly matches the empirical frequencies. 
For all the cases we investigate (with one exception, see Appendix~\ref{app:quality}), we can ensure that the MCMC sampling is done in equilibrium and the resulting PM can be
sampled ergodically in relatively short times, see also~\cite{decelle2021equilibrium} for a detailed discussion.
Note that a PM trained in this way also corresponds to the maximum entropy or least constrained model that is compatible with the one- and two-site empirical frequencies ~\cite{lapedes1999correlated,weigt2009identification}.

Our decimation procedure consists in choosing pairs of sites $i<j$ and amino acids $a,b$, and fixing the corresponding coupling permanently to $J_{ij}(a,b)=0$. The coupling is removed from the set of adjustable parameters, and the corresponding two-site frequency $f_{ij}(a,b)$ is no longer explicitly fitted in the subsequent BML epochs.
However, an important property of PM is the so-called ``gauge'' or reparameterization invariance: the transformation
 \begin{equation}\begin{split}
 J_{ij}(a,b) &\to J_{ij}(a,b) +{\cal J}_{ij} (a) + {\cal K}_{ij} (b) \ , \\
 h_i(a) &\to h_i(a)-{\cal H}_i -\sum_{j (> i)} {\cal J}_{ij} (a)-\sum_{j (< i)} {\cal K}_{ji} (a) \ ,
\end{split} 
\label{eq:gauge}
\end{equation}
leaves the Hamiltonian in Eq.~\eqref{eq:H} and the associated Boltzmann distribution $P(S)$ invariant, for any choice of the $\JJ$, $\KK$ and ${\cal H}$. 
Hence, a gauge transformation can transform a zero coupling into a non-zero one and vice versa.
Because the decimation procedure fixes some couplings to zero, it breaks this invariance. 

\begin{figure}[t]
\begin{center}
\includegraphics[width=\columnwidth]{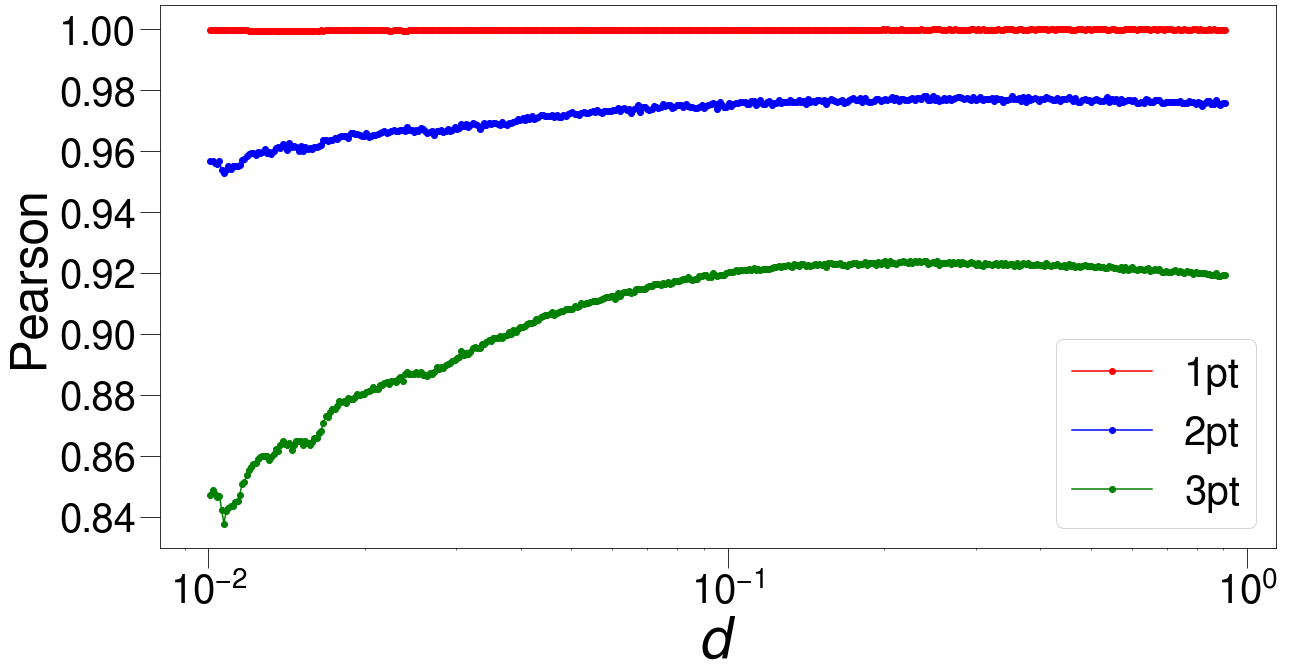} 
\caption{ {\bf Fitting and generative quality for PF00076 --} Pearson correlation coefficient between model and data frequencies as a function of the model density. 
The one-site frequencies $f_i(a)$ are directly fitted. The two-site connected correlations $c_{ij}(a,b)$  are fully fitted by the densest model, while only a fraction of them are fitted for the sparse models at $d<1$. The three-site connected correlations $c_{ijk}(a,b,c)$ are never fitted. The generative performance of the model is essentially unchanged down to a density of 10\%, and slowly decays for even sparser couplings. However, even down to $d=1\%$, the Pearson coefficients remain at remarkably high values above 95\% for the two-site correlations, and above 84\% for the three-site correlations. } 
 \label{fig:pearson}
\end{center}
\end{figure}

We thus begin our decimation procedure by a ``gauge fixing'' step, which sets to zero $2q-1$ out of all $q^2$ entries of each coupling matrix $J_{ij}$. To do so, we identify, independently for each pair of sites $1\leq i<j\leq L$, the $2q-1$ amino-acid pairs $(a,b)$ of smallest connected correlation $c_{ij}(a,b)=f_{ij}(a,b)-f_i(a)f_j(b)$, and fix the corresponding couplings $J_{ij}(a,b)$ to zero. Only the other $q^2-2q+1 = (q-1)^2$ couplings are updated using the BML, Eq.~\eqref{eq:PMlearning}. This procedure chooses a model of minimal density $d = [(q-1)/q]^2 = 90.7\%$ out of all equivalent PM related by the gauge transformation in Eq.~\eqref{eq:gauge}.
The parameters are initialized using a ``profile model'' fitting only the one-site frequencies $f_i(a)$. This initial model has zero couplings and fields $h_i^{(0)}(a) = \log f_i(a) +{\cal H}_i$, with the constant ${\cal H}_i$ being fixed by $\sum_a h_i^{(0)}(a) =0$ (with a very small pseudo-count added to $f_i(a)$ to avoid infinite fields, see Appendix~\ref{app:training}).
The fitting quality of the learned PM is tested by the Pearson correlation between the empirical $c_{ij}(a,b)$ and their counterparts in the model $P(S)$, the latter being estimated from a large independently and identically distributed MCMC sample. For all protein families considered in this work, this Pearson correlation exceeds 0.95, see Fig.~\ref{fig:pearson} and Appendix~\ref{app:quality}.
Note that the results of our decimation procedure depend on the initialization and gauge fixing described above. We tried a different initialization, either fixing both couplings and fields to zero, or initializing both using pseudo-likelihood maximization (PLM) \cite{ekeberg2013improved}. We found qualitatively similar results, but with slightly worse performance (Appendix~\ref{app:init}).

Any further decimation of couplings changes the model. To measure the impact of removing a given coupling $J_{ij}(a,b)$ from a PM, we determine the symmetric Kullback-Leibler (KL) divergence between the Boltzmann distributions with and without that coupling. 
We thus consider a Potts Model with Hamiltonian $H$, and another with Hamiltonian $H'$ in which a given coupling is removed:
\begin{equation}
    H'(S)=H(S) + J_{ij}(a,b) \delta_{a,s_i} \delta_{b,s_j} \ .
\end{equation}
We observe that averages over $P'= e^{-H'}/Z'$ can be expressed in terms of averages over $P= e^{-H}/Z$ as
\begin{equation}\begin{split}
    \langle O(S) \rangle_{P'} 
    &= \frac{\sum_S O(S) e^{-H'(S)}}{\sum_S e^{-H'(S)}}\\
&    = \frac{\sum_S O(S) e^{-J_{ij}(a,b) \delta_{a,s_i} \delta_{b,s_j}} e^{-H(S)}}{\sum_S e^{-J_{ij}(a,b) \delta_{a,s_i} \delta_{b,s_j}}e^{-H(S)}} \\
 &   =\frac{\langle O(S) e^{-J_{ij}(a,b) \delta_{a,s_i} \delta_{b,s_j}} \rangle_P}{\langle e^{-J_{ij}(a,b) \delta_{a,s_i} \delta_{b,s_j}} \rangle_P} \ .
\end{split}\end{equation}
Hence, the symmetric Kullback-Leibler divergence of $P$ and $P'$ is
\begin{eqnarray}
D_{ij}^{ab} &=& D_{\rm KL}(P||P') +D_{\rm KL}(P'||P) \nonumber \\
&=& -\sum_S [P(S)-P'(S)] [\log P(S) - \log P'(S)] \nonumber\\
&=&
\langle H'-H \rangle_P - \langle H'-H \rangle_{P'}\nonumber \\
&=&\langle J_{ij}(a,b) \delta_{a,s_i} \delta_{b,s_j} \rangle_P - \langle J_{ij}(a,b) \delta_{a,s_i} \delta_{b,s_j} \rangle_{P'} \label{eq:DKL}\\
&=&\langle J_{ij}(a,b) \delta_{a,s_i} \delta_{b,s_j} \rangle_P \nonumber\\ 
&& -
\frac{\langle J_{ij}(a,b) \delta_{a,s_i} \delta_{b,s_j} e^{-J_{ij}(a,b) \delta_{a,s_i} \delta_{b,s_j}} \rangle_P}{\langle e^{-J_{ij}(a,b) \delta_{a,s_i} \delta_{b,s_j}} \rangle_P}\nonumber \\
&=& J_{ij}(a,b) p_{ij}(a,b) - \frac{J_{ij}(a,b) p_{ij}(a,b) e^{-J_{ij}(a,b)}}{ p_{ij}(a,b) e^{-J_{ij}(a,b)}+1 - p_{ij}(a,b)} 
\ ,\nonumber
\end{eqnarray}
where $p_{ij}(a,b)=\langle \delta_{a,s_i} \delta_{b,s_j} \rangle_P$ is the marginal two-site probability of $P$, which coincides, at convergence of Eq.~\eqref{eq:PMlearning}, with the empirical frequency $f_{ij}(a,b)$.
Note that we could also have equivalently used the non-symmetrized KL divergence (Appendix~\ref{app:score}).

At each decimation step, we now remove the least significant couplings, {\em i.e.}~those with the lowest $D_{ij}^{ab}$. For computational efficiency, this is done for a fixed fraction (in this work we choose $1\%$) of all remaining couplings.
Note that $D_{ij}^{ab}=D(J,p\sim f)$ (dropping the indices for notational simplicity) goes to $0$ either when $f \to 0$ or $f \to 1$ at fixed $J$, or when $J \to 0$ at fixed $f$. 
More precisely, we have $D(J,f)\sim Jf(1-e^{-J})$ for $f\rightarrow 0$, $D(J,f)\sim J(f-1)$ for $f\rightarrow 1$, and $D(J,f)\sim J^2f(1-f)$ for $J\rightarrow 0$. 
The first and second limits imply that finite couplings can be decimated if the corresponding frequency is close to zero or one,
{\em i.e.}~they have little statistical significance because the corresponding amino acids are almost never observed
(as in color-compression \cite{rizzato2020inference}) or almost always observed. The third limit indicates that small couplings are decimated whatever $f$ is (similar to the procedure proposed in \cite{decelle2014pseudolikelihood} for the inverse Ising problem using PLM).
Numerically, we observe that the percentage of pruned couplings corresponding to each category varies during decimation (Appendix~\ref{app:decimatedcouplings}). After a decimation step, we perform additional BML iterations of Eq.~\eqref{eq:PMlearning} on all undecimated couplings and the fields, to reach convergence again. 
In this way, we progressively obtain PMs of reduced density, and we stop the decimation when $d=1\%$.

Note that in order to accurately estimate $D_{ij}^{ab}$, it is important that the PM learning is well converged before each decimation step. We attempted an ``online'' decimation in which couplings are pruned either after a fixed number of iterations of Eq.~\eqref{eq:PMlearning} or for having reached convergence, and found that this provides no advantage (Appendix~\ref{app:onlinelearning}), neither in terms of generative performance (\ie the Pearson correlations at $d=1\%$ do not improve), 
nor in computational efficiency (\ie the computational time required to reach $d=1\%$ is almost unchanged). Other decimation strategies based on $f_{ij}(a,b)$ only (removing statistically unsupported couplings), or on $J_{ij}(a,b)$ only (removing small couplings), or on applying $\ell_1$-norm regularization to select relevant couplings, were found to perform substantially worse than the information-based procedure using  Eq.~\eqref{eq:DKL} (Appendix~\ref{app:decimationstrategies}).

We have also tested our decimation procedure on synthetic data (see
Appendix~\ref{app:synth}) and found that it is able to correctly identify the
ground-truth sparse model, provided enough data are available.

\section{Results and discussion}
 
We focus here on a representative protein family, the PF00076 family from the Pfam database~\cite{el2019pfam}, corresponding to a RNA recognition motif (RRM) of about 90 amino acids, known to bind single-stranded RNAs. The MSA provided by Pfam contains $M = 137605$ sequences of aligned length $L = 70$. Results obtained for other families (Appendix~\ref{app:otherproteins}) fully confirm the general conclusions drawn here for the RRM. The features of the protein families used for this work are reported in Appendix~\ref{app:data}.

In Fig.~\ref{fig:pearson} we show, for model densities down to 1\%, the Pearson correlation coefficient between the empirical one-site frequencies $f_i(a)$ obtained from the original MSA, and the model one-site marginal probabilities $p_i(a)$, estimated by MCMC sampling.
Similar curves are also shown for the two-site connected correlations $c_{ij}(a,b)$ and for a selected sub-set (specified in Appendix~\ref{app:quality})
of three-site connected correlations, defined as 
\begin{equation}\begin{split}
c_{ijk}(a,b,c) &= f_{ijk}(a,b,c)
    -f_{ij}(a,b)f_k(c) \\ -f_{jk}(b,c)f_i(a) 
    & - f_{ki}(c,a)f_j(b)+2f_i(a)f_j(b)f_k(c) \ , \label{eq:c3}
\end{split}\end{equation}
where $i,j,k$ are the indices of the columns of the MSA (which take value from $1$ to $L$), and
$a,b,c$ run over the amino-acids and the gap symbol (practically, from 1 to $q$). 
The one-site frequencies are perfectly reproduced by the model, {\em i.e.}~$f_i(a)=p_i(a)$, as a consequence of the fixed-point condition in Eq.~\eqref{eq:PMlearning}, and the Pearson coefficient thus remains equal to one at all $d$ (up to tiny deviations due to the finite MCMC samples used in BML and in estimating $p_i(a)$). 
For the maximal density $d_{\rm max}=[(q-1)/q]^2$ obtained after gauge fixing, the two-site correlations should also be perfectly reproduced because of Eq.~\eqref{eq:PMlearning}. 
In practice we only 
reach a Pearson coefficient of $\sim0.975$ due to sampling noise (Appendix~\ref{app:sampling}).
On the contrary, for $d<d_{\rm max}$ only a fraction of all two-site frequencies is explicitly fitted by the model via sparse BML. 
Nevertheless the two-site Pearson coefficient is essentially independent of $d$, up to a slight reduction when $d<10\%$. Finally, three-site correlations, that are never explicitly fitted by the model (the training process in Eq.~\eqref{eq:PMlearning}
does not include three-site information), are nevertheless very accurately reproduced, with a Pearson coefficient around $0.94$ for all $d>10\%$. Note that the reproduction of unfitted observables is a highly non-trivial test for the generative properties of our models~\cite{figliuzzi2018pairwise}, {\em i.e.}~of the capacity of the model to generate data being statistically close to indistinguishable from the natural sequence data used for model learning.  Below density $d=10\%$, the generative quality
of the model for three-site correlations is slightly reduced, remaining nevertheless very high (above 84\% down to $d=1\%$). We discuss the generative property of the sparse models introducing additional metrics in Appendix~\ref{app:similarity}.

\begin{figure}[t]
\begin{center}
\includegraphics[width=\columnwidth]{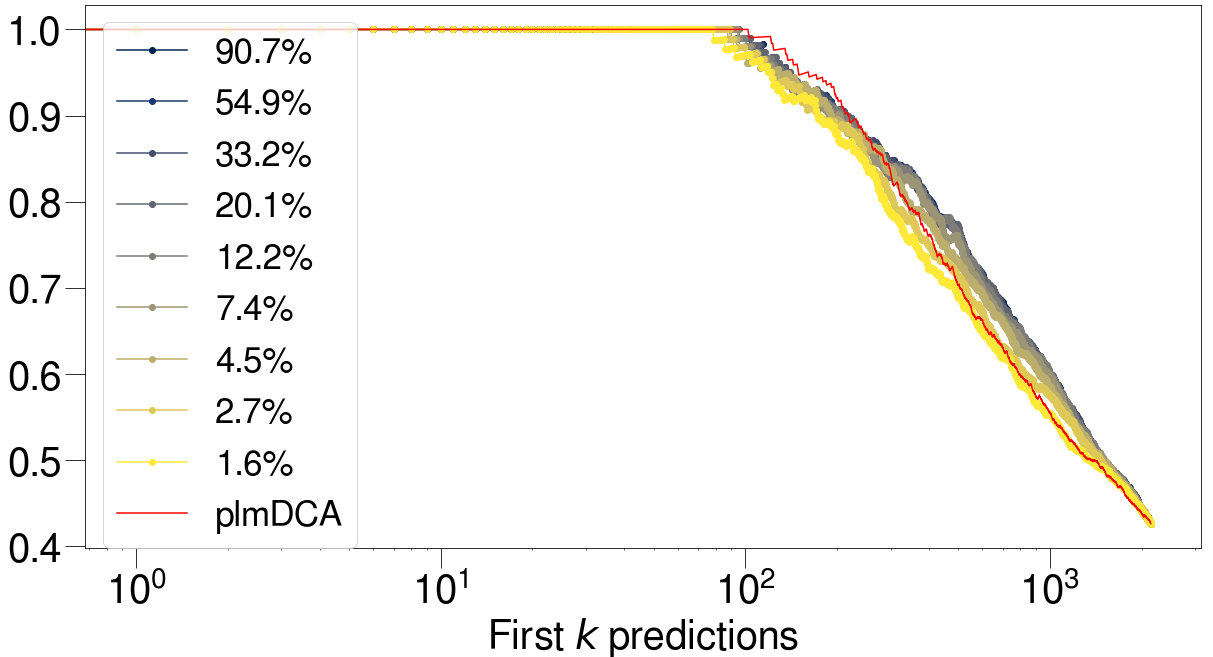} 
\caption{ {\bf Contact prediction for PF00076 --} Positive predictive values (PPV) for several model densities, {\em i.e.}~the fraction of true positives
among the highest-ranking $k$ pairs $(i,j)$ of sites, when ordered by decreasing $F_{ij}^{\rm APC}$. Even the most sparse model, with only 
1.6\% of couplings, shows an excellent performance at contact prediction. The curve for plmDCA, a standard DCA approach for contact prediction, is shown for reference and gives comparable results.} 
 \label{fig:PPV}
\end{center}
\end{figure}

A second test of model quality is the prediction of structural contacts, which constituted the major application of DCA in the last years. The idea is that pairs of strongly interacting sites in the PM should
correspond to close-by residues in the three-dimensional structure, which display strong coevolution to maintain the proper protein fold and functionality. Using the standard convention for coevolutionary contact prediction, we consider a pair of residues to be in contact if the distance between them is at most 8~$\mathring{\mathrm{A}}$, and we exclude easy-to-predict short-range contacts by considering only pairs with $|i-j|\geq 4$ in our analysis.
The reference (ground-truth) distance was obtained by the package~\cite{edoardo_sarti_andrea_pagnani_2020} that takes the shortest distance between heavy atoms in all protein structures registered in the Protein Data Bank (PDB)~\cite{berman2007worldwide} for the given Pfam family. 
We follow the standard procedure for contact prediction, which consists in computing the average-product corrected (APC) Frobenius norms of the coupling matrices
(note that the coupling matrices are transformed into the zero-sum gauge and that the gap states $a,b=q$ are excluded from the sum \cite{feinauer2014improving}),
\beq
F_{ij} = \sqrt{ \sum_{a,b=1}^{q-1} J_{ij}(a,b)^2 }\ ;
\ F_{ij}^{\rm APC} = F_{ij} - \frac{\sum_k F_{ik} \sum_k F_{k j}}{\sum_{kl} F_{kl}} \ .
\eeq
In Fig.~\ref{fig:PPV} we show the fraction of true contacts within the first $k$ pairs of sites, ranked in decreasing order of $F_{ij}^{\rm APC}$. We observe that the performance of the model at inferring the structural contacts is only slightly deteriorated even in the sparsest case $d=1.6\%$.

\begin{figure}[t]
\begin{center}
\includegraphics[width=0.95\columnwidth]{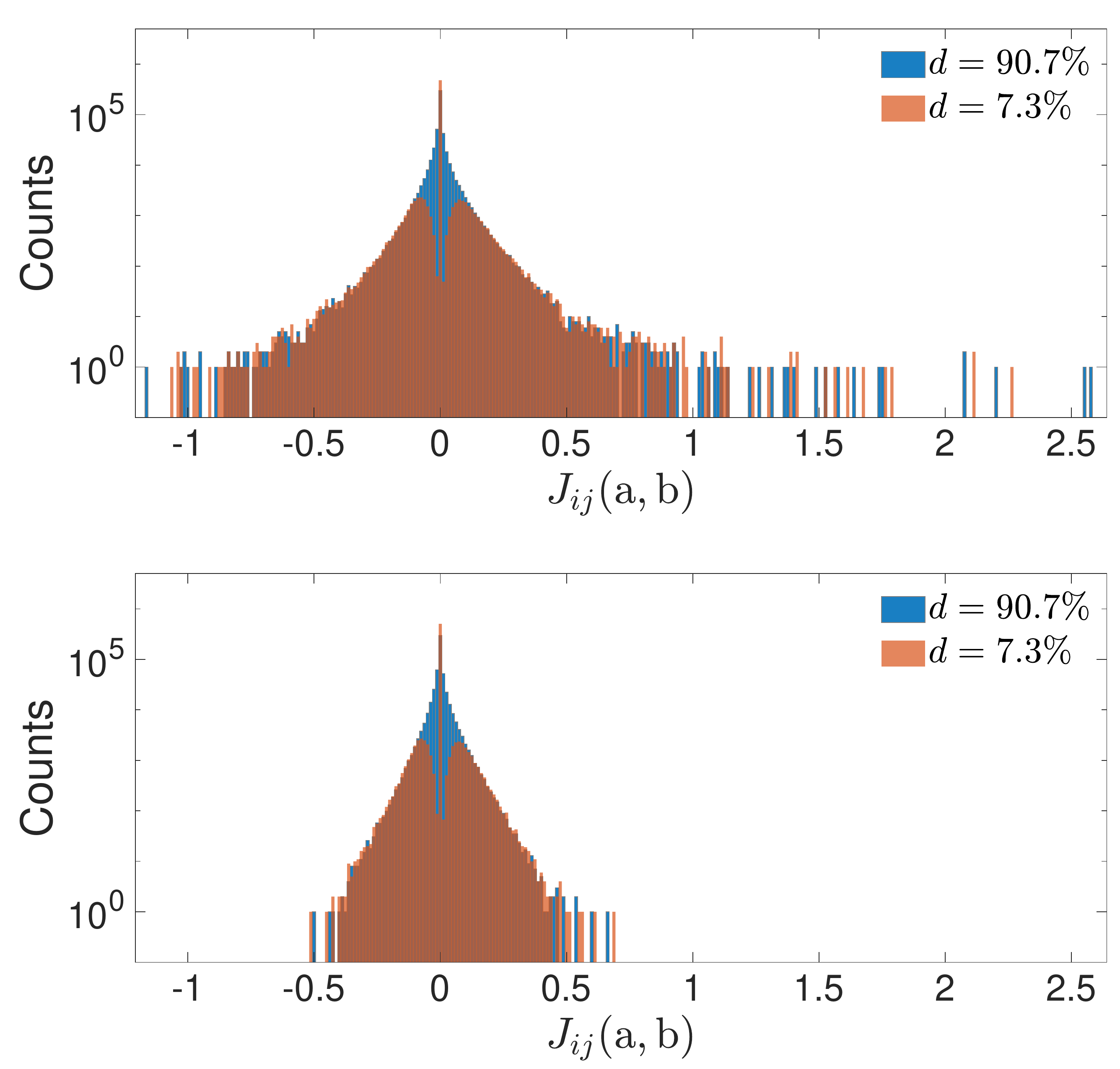} 
\caption{ {\bf Coupling distributions --} Distribution of couplings corresponding to true 
contacts (top) and to non-contacts (bottom) in the three-dimensional protein fold, for the initial PM with density $d_{\rm max}=91\%$ and a sparser model having density $d=7\%$ associated with a reasonably accurate contact prediction. 
} 
 \label{fig:interpretation}
\end{center}
\end{figure}

In Fig.~\ref{fig:interpretation} we show the probability distributions of couplings $J_{ij}(a,b)$, separately for pairs $i<j$ corresponding to contacts and all the others. We observe that, both for contacts and non-contacts, the decimation affects the shape of the distribution around $J\sim 0$ in a similar way, while the tails are essentially unaffected.
Overall, these
results explain why the performance of the PM for contact prediction using $F_{ij}^{\rm APC}$ 
is essentially independent of $d$ (Fig.~\ref{fig:PPV}). Unfortunately, the large-$J$ tail of the distributions of 
couplings on non-contacting sites does not change upon sparsifying the model, which suggest that our decimation procedure cannot help in devising better contact predictors.

\begin{figure}[t]
\begin{center}
\includegraphics[width=\columnwidth]{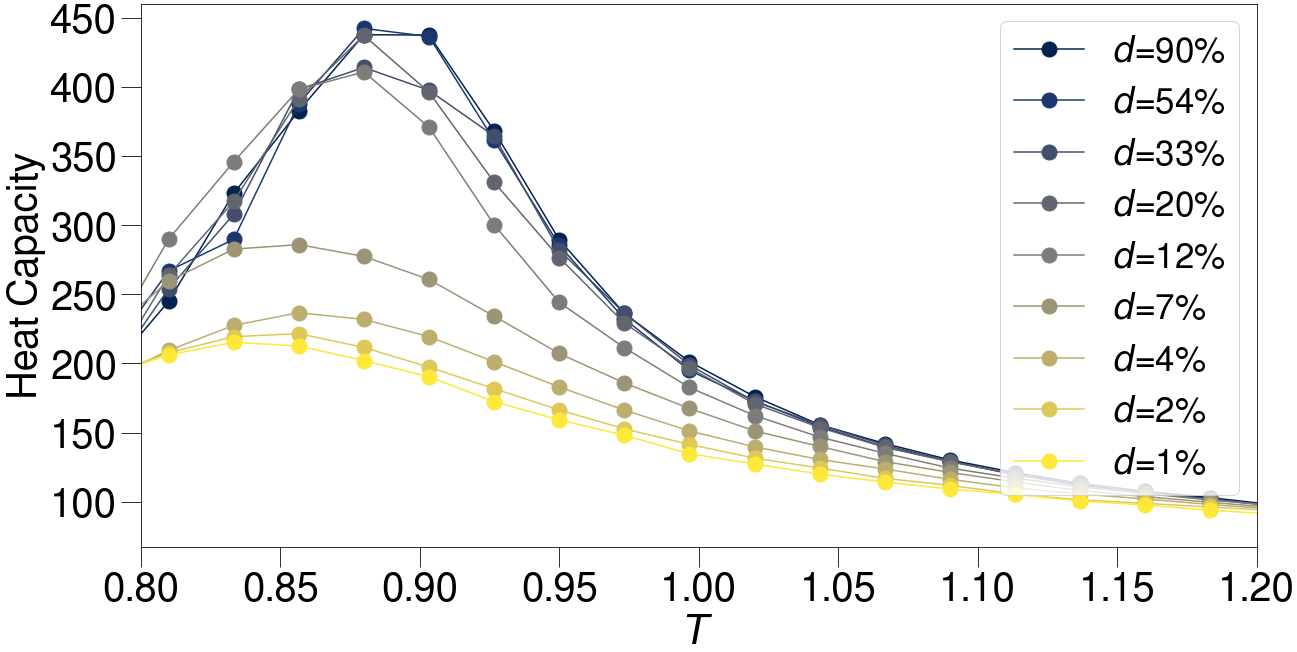} 
\caption{ {\bf Criticality --} Heat capacity as a function of temperature for models with different density. The densest models show
a strong peak of specific heat close to the reference scale $T=1$, which is a signature of criticality: the model is extremely sensitive to a small change of couplings, due
to overfitting. On the contrary, sparse models display a much smaller peak, which is also shifted away from $T=1$ towards lower temperatures, indicating a better 
robustness of the learning.
} 
 \label{fig:CV}
\end{center}
\end{figure}

In order to study the criticality of the models, we consider a simple perturbation of the statistical weight, 
{\em i.e.}~we rescale the Hamiltonian $H(S)$ by a formal inverse
temperature $\b = 1/T$ and set $P(S)\sim e^{-\b H(S)}$, in such a way that $T=1$ corresponds to the original model trained on data, while measuring the variation of the model entropy $S$.
In Fig.~\ref{fig:CV} we show the heat capacity $C = T \de S/ \de T$ of the PM for several sparsities (see Appendix~\ref{app:criticality} for details on the computation of $C$).
Note that a large $C$ indicates a large variation of the model entropy with $T$, or equivalently that the model 
statistics changes strongly after a slight change of the parameters. This is indeed 
the best definition of criticality in statistical physics, keeping in mind that our models have finite size $L$ and we thus cannot perform a finite-size scaling analysis
to determine if the observed peak in $C$ corresponds to a phase transition in the thermodynamic limit.
In Fig.~\ref{fig:CV} we observe that the denser models display a large peak in $C$ close to $T=1$, which indicates that the models are close to criticality.
Upon sparsifying the model, the peak amplitude is strongly reduced and the peak is also shifted to lower temperatures, {\em i.e.}~further away from the reference scale $T=1$.
These results suggest that the criticality of the dense models comes from over-fitting. Because the dense models have a huge number of parameters, they
are able to fit all the details of the training data. As a consequence, the model becomes very sensitive to noise, and a little change of the parameters changes a lot the model
statistics. On the contrary, sparse models have less parameters and are thus more robust to noise.

\begin{figure}[t]
\begin{center}
\includegraphics[width=\columnwidth]{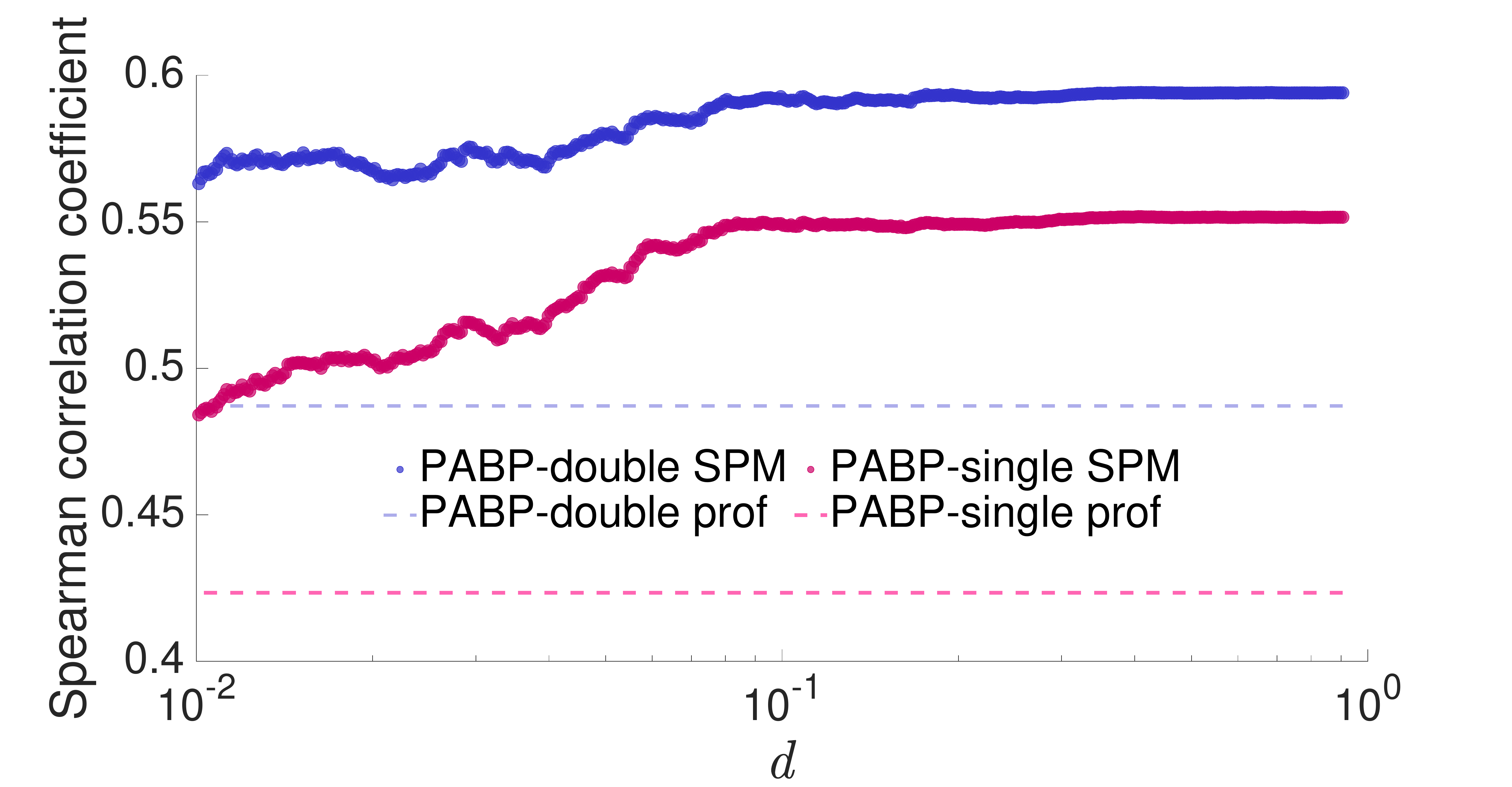} 
\caption{ {\bf Single and double mutations --} 
Spearman correlation between the experimental fitnesses and the model predictions as a function of the model density,
both for single and double mutants of the PABP, a member of the PF00076 family. The dashed lines show the same correlations for a profile model ({\em i.e.}~$d=0$) as a reference.
} 
 \label{fig:mut}
\end{center}
\end{figure}

Ref.~\cite{melamed2013deep} provides Deep Mutational Scanning (DMS) data for a representative member of the PF00076 family, namely the RRM2 domain of the Poly(A)-Binding Protein (PABP) in the yeast species {\it Saccharomyces cerevisiae}. Using this domain as a reference,
the authors generated a library of 110,745 protein variants, including 1,246 single amino-acid substitutions 
and 39,912 double amino-acid substitutions. Each of these variants was experimentally scored for function, by monitoring the growth of mutant yeast and finally, a ``fitness score'' was attributed to each mutant sequence in the experiment~\cite{melamed2013deep}. Within our models, the inferred Hamiltonian $H(S)$ in Eq.~\eqref{eq:H}
is also interpreted as a sequence-fitness score. Hence, a good test of the generative property of our models is to check whether the energy differences $\Delta H = H({\rm mutant})- H({\rm reference})$ between mutant sequences and the PABP reference
correlate well with the experimental fitness variations.  Because the mapping between experimental and model fitness may be non-linear, in Fig.~\ref{fig:mut} we show the Spearman's rank correlation between these two variables, both for single and double mutants. In the dense $d=d_{\rm max}$ case, we reproduce the reference values already given in Ref.~\cite{hopf2017mutation}. We also observe 
that upon reducing density, once again the model quality is not degraded, down to $d\sim 10\%$. Even for $d=1.6\%$ the model performs quite well, and much better than a profile model, which coincides with the limit $d\to 0$ of our decimation procedure.

\section{Conclusions}
 
We introduced a general parameter reduction scheme for Boltzmann Machine Learning, and we applied it to Potts models for protein sequence data, {\em i.e.}~for the learning of highly accurate and generative, but sparse DCA models. Our strategy makes use of a rigorous information-based criterion to select couplings that are iteratively pruned. Intuitively, removed couplings are either statistically unsupported, {\em i.e.}~they correspond to pairs of amino acids that are almost never or almost always observed, similarly
to the color-compression scheme~\cite{rizzato2020inference},  or they are small, {\em i.e.}~they correspond to pairs that are only weakly correlated, or a combination of both. The statistical significance of a coupling is precisely quantified by the symmetric KL divergence between the Boltzmann measure of the Potts model with and without this coupling, which 
is exactly computable from the model or the empirical statistics.

While our method is fully general for learning Boltzmann machines from high-dimensional categorical data, here we focused on its application to model protein families via Potts models, in which strong couplings are usually associated with physical contacts in the three-dimensional protein fold. We stress that the aim of this work is not to provide a sparse graph topology underlying the true interaction network, and indeed the pruning is not performed block-wise but at the level of the individual coupling entries, but to provide a general framework of parameters reduction strongly based on information-theoretic assumptions.
We have shown that the model can be decimated down to less than $10\%$ of the original couplings, while losing neither its generative quality, 
nor its accuracy in contact prediction.
However, it has to be noted that many couplings not corresponding to structural contacts remain non-zero even in the lowest-density models. The interpretation of such couplings remains unclear. They may result from subtle effects due to the phylogenetic relations between the training sequences \cite{qin2018power,horta2020phylogenetic}, but also from extended functional constraints as those found by Restricted Boltzmann Machines or Hopfield-Potts models~\cite{tubiana2019learning,shimagaki2019selection}. As a result, further work is needed to make DCA-type modeling fully interpretable.

The sparse models resulting from our decimation procedure are also far away from criticality: they do not display the specific-heat peak close to the formal temperature $T=1$ that characterizes the dense models. Hence, we attribute the criticality observed in dense models to over-fitting, and conclude that our decimation procedure makes model learning more robust to finite-sample noise. Finally, the model maintains its performance in predicting the fitness of mutations around a reference sequence, {\em i.e.}~it is capable of predicting the {\it local} shape of the fitness landscape after having been trained on a {\it global} alignment of distantly diverged amino-acid sequences.

Our decimation procedure solves the first two problems mentioned in the introduction: we can eliminate small and noisy couplings, and the resulting model maintains its
fitting and generative qualities, while being statistically more robust. Unfortunately, we were unable to solve the third problem, namely the strong dependence on the initial condition of the training: different initial conditions (zero couplings and fields, profile model, plmDCA) produce fully-connected PMs of equal fitting quality but with slightly different performances in predicting contacts and mutational effects. This difference does not disappear after decimation (Appendix~\ref{app:init}). In other words, our decimation procedure remains sensitive to the initial fully-connected model from which it is started.

The resulting sparse PMs attempt to fit the data by using the minimal number of two-site couplings, {\em i.e}~using coupling matrices that are as sparse as possible. In the context of proteins (or RNA), it is natural to think that the sparse couplings identified by the model are the most relevant to describe the physical two-site correlations
that arise from the need to maintain the three-dimensional folded structure.
This strategy is complementary to collective-feature learning, {\em e.g.}~via RBM or Hopfield-Potts models~\cite{tubiana2019learning,shimagaki2019selection}, in which the number of parameters in the coupling matrix is reduced by
assuming it to be of low-rank. The features learned by these machines are associated with global sequence motifs, related, {\em e.g.}, to protein function or its interactions, but the accuracy of contact prediction is reduced.
An interesting and natural perspective would be to combine these two strategies into a general ``sparse plus low-rank'' scheme, \cf~\cite{zhang2016improving} for a related idea,
which could lead to an accurate description of protein families in an easily interpretable way, with sparse two-site couplings describing
physical constraints coming from structural contacts, and low-rank couplings describing biological features associated with protein function and its evolutionary history. 

To conclude,
we would like to stress once more that our information-based decimation strategy is not specific to the application of Potts models to protein sequence data. 
It can directly be used in other applications of inverse statistical physics and Boltzmann machine learning, 
as \eg in modeling neural or socio-economic data~\cite{nguyen2017inverse}, and may be adapted to more general network reconstruction schemes.

The code for learning and pruning the PMs is available at ~\cite{adabmDCA}. 

\acknowledgments

  We would like to thank Matteo Bisardi, Simona Cocco, Yaakov Kleeorin, R\'emi Monasson, Rama Ranghanatan, Olivier Rivoire and Jeanne Trinquier for discussions related to this work.
  We acknowledge funding by the EU H2020 research and innovation programme MSCA-RISE-2016 (grant 734439 InferNet, to MW), by the Simons Foundation (\#454955, to FZ) and by the Honjo International Scholarship Foundation (PhD grant to KS).

P.B.-C., A.P.M. and K.S. contributed equally to this work. Author contributions:
P.B.-C. and M.W. designed the research;
P.B.-C., A.P.M., K.S., and F.Z. performed the research;
A.P.M. and K.S. analyzed the data;
A.P.M., K.S., M.W., and F.Z. wrote the paper.

\appendix

\section{Methods}
\label{sec:MCMC}

\subsection{Alternative decimation score}
\label{app:score}

Using the relation
\begin{equation}\begin{split}
    \frac{Z'}{Z} &= \frac1Z \sum_S e^{-J_{ij}(a,b) \delta_{a,s_i} \delta_{b,s_j}} e^{-H(S)} \\
    &= \langle e^{-J_{ij}(a,b) \delta_{a,s_i} \delta_{b,s_j}} \rangle_P \\
    &= p_{ij}(a,b) e^{-J_{ij}(a,b)}+1 - p_{ij}(a,b) \ ,
\end{split}\end{equation}
we obtain
\begin{equation}\label{eq:DKLnsym}\begin{split}
    \widehat D_{ij}^{ab} &= D_{\rm KL}(P||P')=\sum_S P(S)[\log P(S) - \log P'(S)] \\
    &= J_{ij}(a,b) p_{ij}(a,b) \\&+ \log[ p_{ij}(a,b) e^{-J_{ij}(a,b)}+1 - p_{ij}(a,b)] \ .
\end{split}\end{equation}
This second quantity also coincides with the variation of the likelihood of data under the change of model,
\begin{equation}\begin{split}
    \Delta {\cal L} &= \frac1M \sum_{m=1}^M [\log P(S_m) - \log P'(S_m)] \\
    &= \frac1M \sum_{m=1}^M J_{ij}(a,b) \delta_{a,s^m_i} \delta_{b,s^m_j} + \log \frac{Z'}{Z} \\& =
    J_{ij}(a,b) f_{ij}(a,b) \\&+ \log[ p_{ij}(a,b) e^{-J_{ij}(a,b)}+1 - p_{ij}(a,b)] \ ,
\end{split}\end{equation}
which coincides with~\eqref{eq:DKLnsym} when the model is well converged and $p_{ij}(a,b)=f_{ij}(a,b)$.

Note that the qualitative form of $D_{ij}^{ab}$ and 
$\widehat D_{ij}^{ab}$ as a function of $f_{ij}(a,b)$ and $J_{ij}(a,b)$ is very similar, and $D_{ij}^{ab}$ is a monotonous function of $\widehat D_{ij}^{ab}$. Using 
$D_{ij}^{ab}$ or $\widehat D_{ij}^{ab}$ in the decimation procedure thus leads to fully equivalent results.

\subsection{Test on synthetic data}

\label{app:synth}
To evaluate the accuracy of our information-based decimation procedure, we learn and prune a fully connected model learned from a set of synthetically generated sequences sampled from a known sparse model (a ground-truth), whose parameters will be compared to our results. The true model is a Viana-Bray model of 50 Ising spins lying in a random regular graph of degree 4. The couplings associated with the 100 links are drawn from a Gaussian distribution of zero mean and unit variance. We sample $M$ independent configurations, with $M = \{200, 500, 1000, 5000, 10000\}$, from the associated Boltzmann distribution at $\beta = 0.6$; this value guarantees the sampling to be performed in the paramagnetic phase but close to the phase transition, expected at $\beta_{c} \sim 0.7$ for the same model in the thermodynamic limit \cite{mezard_bethe_2001} (although for this instance the number of variables is finite, we assume the critical point to be closer to that found at the thermodynamic limit). 

\begin{figure*}[t]
	\centering
	\includegraphics[width=0.75\linewidth]{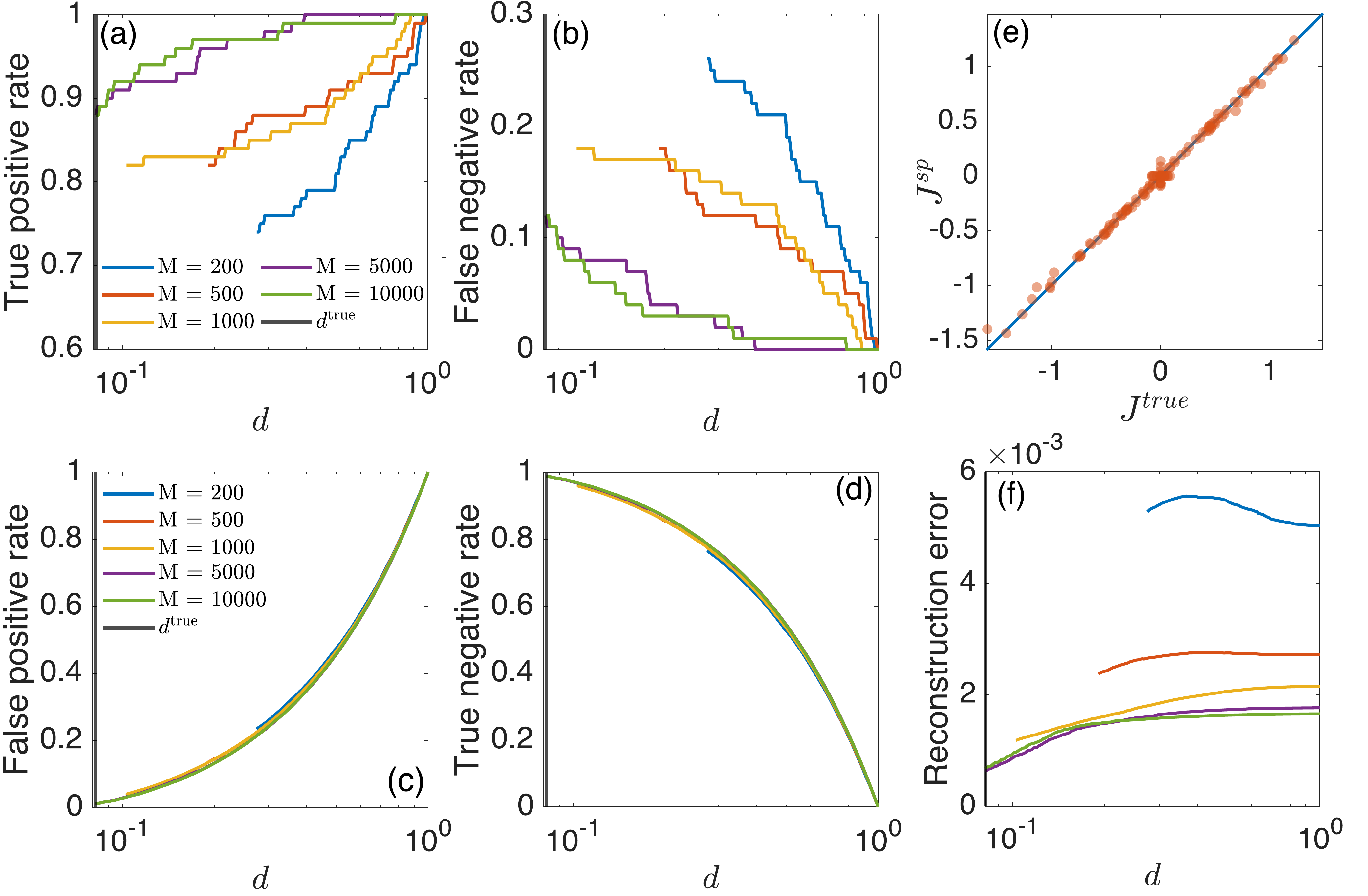}
	\caption{Reconstruction performances measured by the true positive rate (a), false negative rate (b), false positive rate (c), true negative rate (d) and $\ell_{2}$-norm between the true and inferred parameters (f), as a function of the density for the controlled experiment over a Viana-Bray model. We use different colors according to the value of $M$, the number of sequences used within the learning step of the fully connected and of the sparse model. The vertical line represents the density of the true model $d_{true} = 0.0816$. In (e) we show the scatter plot of the true couplings against the parameters learned by the $M = 10000$ run for $d = d_{true}$.  \label{fig:Ising}}
\end{figure*}

The pruning protocol is applied to each of the dense models decimating $1\%$ of the non-pruned couplings every time the convergence error reaches the threshold of $\varepsilon = 2.5 \times 10^{-2}$; the algorithm stops either because we reach the density of the true model $d_{true} = 0.0816$ or because we reach the maximum number of iterations set for this experiment to $20 000$. We show in Fig.~\ref{fig:Ising}(f) the reconstruction error (computed as the $\ell_{2}-$ norm between the learned and the true couplings) and the true positive, false negative, false positive and true negative rates in Fig.~\ref{fig:Ising}(a), (b), (c) and (d) respectively. Different lines correspond to different $M$ as specified in the legend. For small values of $M$ the decimation procedure is very slow because every time a new pruning step is performed, several learning iterations are needed to reach a new convergence. This explains why in Fig.~\ref{fig:Ising}(a), (b), (c), (d) and (f) the lines associated with $M = \{200, 500, 1000\}$ do not reach $d_{true}$ within the $20 000$ iterations. The reconstruction performance is overall accurate as shown by the reconstruction error although $\sim 20$ of the true couplings are set to zero within the decimation procedure (as stressed by the values of the true positive and false negative rates in (a) and (b) panels). For large $M$, the reconstruction performance (in terms of both the $\ell_{2}$-norm and the number of wrongly/correctly pruned couplings) significantly improves: the algorithm is now able to reach the desired sparsity and the true positive rate is close or above $0.9$ for all densities (only 12 parameters are inaccurately estimated). These couplings have a true value close to zero as shown in Fig.~\ref{fig:Ising}(e) by a scatter plot of the true parameters versus the learned couplings of the $M = 10000$ samples run for $d = d_{true}$. The points forming a cross in the origin of the axes are associated with the 12 zero couplings of the learned model (in correspondence of the 12 non-zeros parameters of the true one) and, similarly, with the 12 non-zero learned parameters that are not present in the true model. 

\subsection{Data set}
\label{app:data}

In the following, we report the details of the five protein families analyzed in our work, identified as PF00014, PF00072, PF00076, PF00595, and PF13354 in the Pfam database (\url{https://pfam.xfam.org/}) \cite{bateman2002pfam, el2019pfam}.
For PF00014, PF00072, PF00076 and PF00595 we filter the full set of sequences downloaded from Pfam, keeping only those that have less than six consecutive gaps. Empirically, we have found that the presence of stretched gaps renders the training more difficult as the Markov Chain Monte Carlo (MCMC) used for sampling has difficulties in visiting both very gapped sequences and gap-free region of the sequence landscape in a proper way, i.e. proportionally to the correct Boltzmann weight. This leads to a systematic bias in the model statistics.
For the Beta-lactamase2 family PF13354, we used a slightly different procedure: we downloaded the Pfam pHMM model for that family, and we scanned the NCBI database to obtain aligned sequences. We then filtered sequences according to two criteria:
(i) 80\% sequence coverage (i.e. less than 20\% gaps) and (ii) redundancy reduction at 80\% 
(so $M_{\rm eff} \approx M$ in this case). We also removed the sequence TEM-1 (which is used as reference
in the deep mutational scanning, as discussed below), and all sequences very similar to it.
Note that because there are overlapping Beta-lactamase families in Pfam, 
our procedure, based on a single pHMM, gives also sequences that would align better to some other family in Pfam, in particular to the Beta-lactamase family PF00144.

In Table~\ref{tab:info} we show the name of the protein domain associated with each family, the length, i.e. the number of columns $L$ of the multiple sequence alignment (MSA), the number of sequences $M$ of the original MSA and $M_{\rm eff}$, the number of statistically relevant sequences after a standard re-weighting of close-by sequences~\citep{morcos2011direct}.

\begin{table*}[t]
\centering
\begin{tabular}{c|c|c|c|c|c}
Identifier & PF00014 & PF00072 & PF00076 & PF00595 & PF13354\tabularnewline
\hline 
Protein domain & Kunitz domain & Response regulator receiver domain & RNA recognition motif & PDZ domain & Beta-lactamase2\tabularnewline
\hline 
$L$ & 53 & 112 & 70 & 82 & 202\tabularnewline
\hline 
$M$ & 13600 & 823798 & 137605 & 36690 & 7515\tabularnewline
\hline 
$M_{{\rm eff}}$ & 4364 & 229585 & 27785 & 3961 & 7454\tabularnewline
\end{tabular}

\caption{ We show here the Pfam identifier, the name of the protein domain, the length, the number of sequences and the effective number of sequences for the families analyzed in our work. \label{tab:info} }

\end{table*}

\subsection{Training protocol} 
\label{app:training}

We specify here the details of the Boltzmann learning used to train the dense Potts model and to refine the non-zero parameters within the decimation run.

First, we compute the data statistics from the input MSA as 
\begin{eqnarray}
f_{i}(a) & = & (1-  \alpha) f_{i}^{\rm emp}(a) + \frac{\alpha}{q} \ , \\
f_{ij}(a, b) & = & (1 - \alpha) f_{ij}^{\rm emp}(a, b) + \frac{\alpha}{q^{2}} \ ,
\end{eqnarray}
with $f_{i}^{\rm emp}(a)$ and $f_{ij}^{\rm emp}(a, b)$ being the one- and two-site frequencies computed from the MSA (for all positions $i,j$ and amino acids $a,b$), and with $\alpha$ being a pseudo-count~\citep{PhysRevE.90.012132} introduced to avoid divergent fields and couplings associated with poor statistics. Here we set $\alpha=1/M_{\rm eff}$ except for the PF13354 family for which we
set $\alpha=10^{-50}$ (we observed that other values of the pseudo-count do not lead to a significant change of the trained models).
Then, we start from a profile model, i.e. all couplings are set to zero and the fields are equal to $h_i(a) = \log[f_i(a)] + {\cal H}_{i}$ with ${\cal H}_{i}$ a constant ensuring $\sum_a h_i(a)=0$. Subsequently we iteratively refine the parameters according to Eq.~\eqref{eq:PMlearning}, using as learning rate $\eta_{J} = \eta_{h} = 5 \cdot 10^{-2} $. We stop the algorithm when the convergence error $\epsilon$, computed as the maximum error attained in the fitting of the two-site connected correlations,
\begin{equation}
\epsilon  =  \max_{i,j,a,b} | f_{ij}(a,b) -f_{i}(a) f_{j}(b) - p_{ij}(a,b) + p_{i}(a)p_{j}(b) | \ ,
\end{equation}
reaches $10^{-2}$ (this value may slightly change depending on the family, up to $5 \cdot 10^{-2}$ for the PF13354 family which is the most difficult to train). At each iteration, we use Metropolis-Hasting MCMC to compute the model statistics $p_i(a)$ and $p_{ij}(a,b)$. We run $N_{\rm chain}$ independent MC chains, with $N_{\rm chain}=3000$ for PF00014 and PF00595, $N_{\rm chain}=1000$ for the longer PF13354, and $N_{\rm chain}=5000$ for the copious families PF00072 and PF00076. 
The chains are initialized at the first iteration from a uniform independent random
distribution over all possible amino-acids, gap included, and are then persistent over iterations, i.e. at each new iteration the chain is initialized from the last configuration of the previous iteration.
Each chain runs for $T_{\rm eq}=20$ MC sweeps (one sweep corresponds here to $L$ single-site Metropolis-Hastings MC steps) before starting to
sample $10$ configurations spaced by $T_{\rm wait}=10$ MC sweeps. Hence, the total number of generated samples in a single iteration is
$10 \times N_{\rm chains}$ (from $10^4$ to $5 \cdot 10^{4}$ depending on the family), and each chain is evolved by 110 MC sweeps in a single iteration.

\subsection{Sampling protocol} 
\label{app:sampling}

Once the training is complete, for the final set of parameters of the Potts Model, we need to generate a new sample, from which we compute the model statistics to be compared with the MSA statistics.
As in training,
the MCMC method used for the sampling is the standard Metropolis-Hasting algorithm, using $N_{\rm chain}$ independent MC chains, initialized from a uniform independent random
distribution over all possible amino-acids, gap included.
Each chain is evolved for $T_{\rm eq}$ MC sweeps in order to achieve equilibration, before we start collecting samples, the waiting time between each sampled configurations being $T_{\rm wait}$ MC sweeps.
We specify in Table~\ref{tab:Sampling} the values
of $N_{\rm chain}$, $T_{\rm eq}$ and $T_{\rm wait}$ and of the total number of collected samples, $M_{\rm MC}$. Note that the conditions for the sampling are different from those used in the learning.

\begin{table}[b]
\centering
\begin{tabular}{c|c|c|c|c|c}
Identifier & PF00014 & PF00072 & PF00076 & PF00595 & PF13354\tabularnewline
\hline 
$M_{{\rm MC}}$ & 30000 & 30000 & 30000 & 30000 & 30000\tabularnewline
\hline 
$T_{{\rm wait}}$ & 60 & 80 & 60 & 90 & 100\tabularnewline
\hline 
$T_{{\rm eq}}$ & 10000 & 50000 & 30000 & 50000 & 50000\tabularnewline
\hline 
$N_{{\rm chain}}$ & 100 & 100 & 100 & 100 & 300\tabularnewline
\end{tabular}
\caption{ We report here the details of the MC sampling performed to evaluate the model statistics.\label{tab:Sampling}}
\end{table}

\begin{figure*}[t]
\centering
 \makebox[\textwidth]{\includegraphics[width=1.00\linewidth]{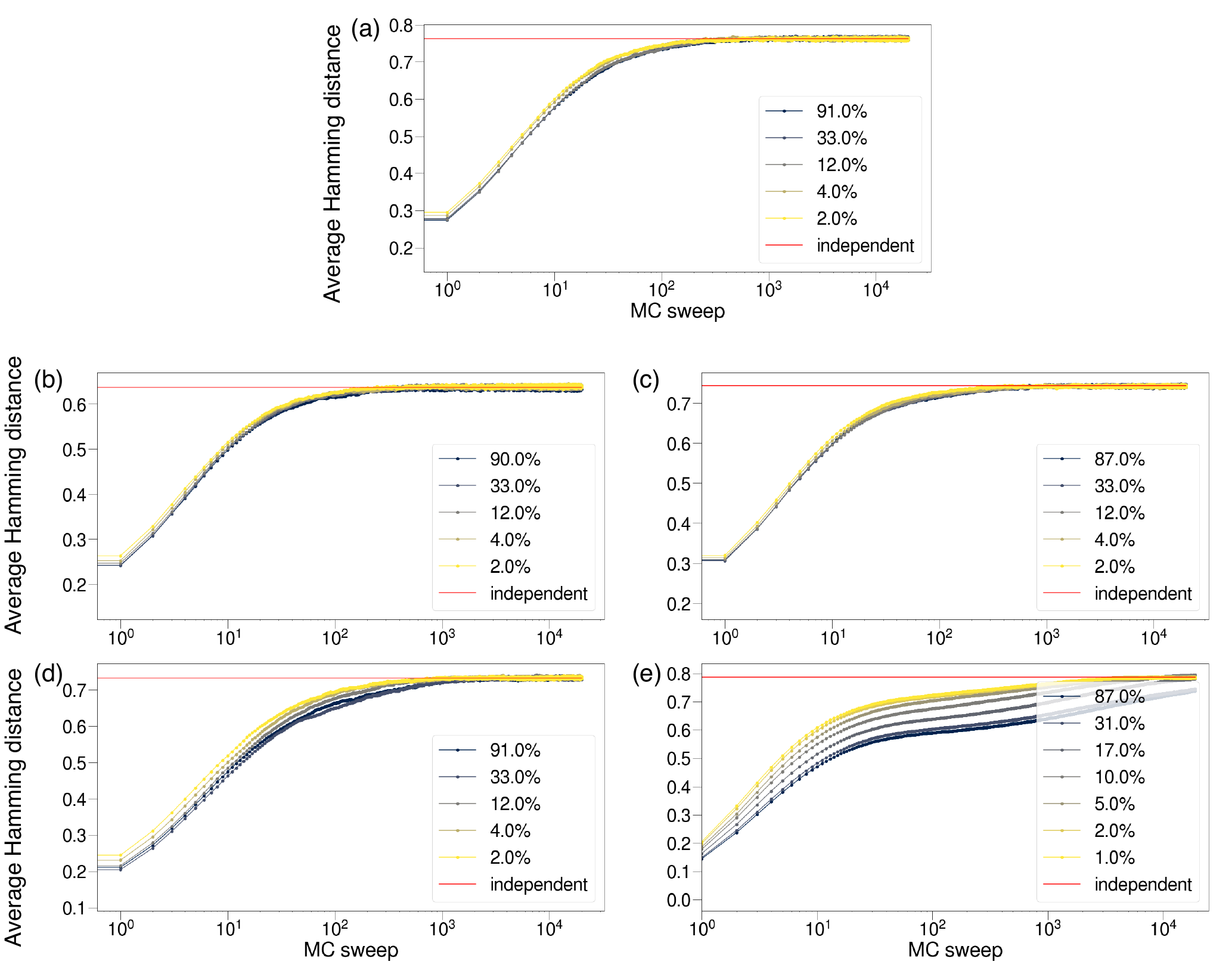}}
\caption{Averaged Hamming distances between an equilibrium sequence at time $t = 0$ and the evolved sequence after $t$ MC sweeps for (a) PF00076 (b) PF00014 (c) PF00072 (d) PF00595 and (e) PF13354. The average is computed using $10^4$ independent MC chains.}  
\label{fig:Hamming_Distance}
\end{figure*}

We also compute, for each model, the Hamming distance $d_H(t)$ between an equilibrium configuration and its time evolution under the MCMC dynamics after $t$ MC sweeps (averaged over initial configurations and over the dynamics), see Fig.~\ref{fig:Hamming_Distance}. Obviously, $d_H(0)=0$ and for short times, $d_H(t)$ grows linearly, with a coefficient given by the acceptance rate of single-site mutations in the MCMC dynamics. At long times, $d_H(t\to\infty)$ saturates at the average distance between two independent samples from the Potts Model equilibrium distribution. This quantity can also be computed by measuring the Hamming distance between two independent MC chains, after equilibration, and is reported as a red horizontal line in Fig.~\ref{fig:Hamming_Distance}.
The time it takes for $d_H(t)$ to reach its asymptotic value gives an estimate of the decorrelation time of the MCMC dynamics, i.e. the time needed to generate a new independent equilibrium sample.

We observe that for PF00014, PF00072 and PF00076, the decorrelation time is of the order of $10^2$ MC sweeps,
and independent of sparsity, which suggest that the model is sampled in equilibrium during the learning process. In fact, we obtain exactly the same model statistics upon resampling the model in different conditions.

For PF00595 the decorrelation time is $\approx 10^3$ MC sweeps for the dense model. Because our training is done with persistent chains, and a small learning rate, we still believe that proper equilibrium sampling is achieved during learning. This is confirmed by the fact that we reproduce the same model statistics under resampling. Furthermore, we observe that the decorrelation time is reduced upon sparsifying the model, which suggests that the sparse models are less critical, as we discuss below.

For PF13354 the situation is radically different. In this case, the decorrelation time is huge (more than $10^4$ MC sweeps for the dense model). This is likely due to the presence of multiple subfamilies, such that the MC chains take a lot of time to jump from one subfamily to another. With such a long decorrelation time, learning becomes extremely hard and we cannot guarantee that equilibration is achieved during it. In fact, we find that upon resampling the model starting from random initial states, the statistics is initially good (after $\approx 2 \times 10^4$ MC sweeps) but then is degraded, indicating that the model suffers from overfitting due to poor equilibration during learning. For the sparse models, the decorrelation time is substantially reduced (by almost a factor 100), and consistently we find that resampling is stable at all times.

\section{Results for the other protein families}
\label{app:otherproteins}

In this section we report the same type of results shown in the main text for PF00076, but for the four remaining families: PF00014, PF00072, PF00595 and PF13354. 

\subsection{Fitting quality}
\label{app:quality}

\begin{figure*}[t]
\centering
 \makebox[\textwidth]{\includegraphics[width=1.00\linewidth]{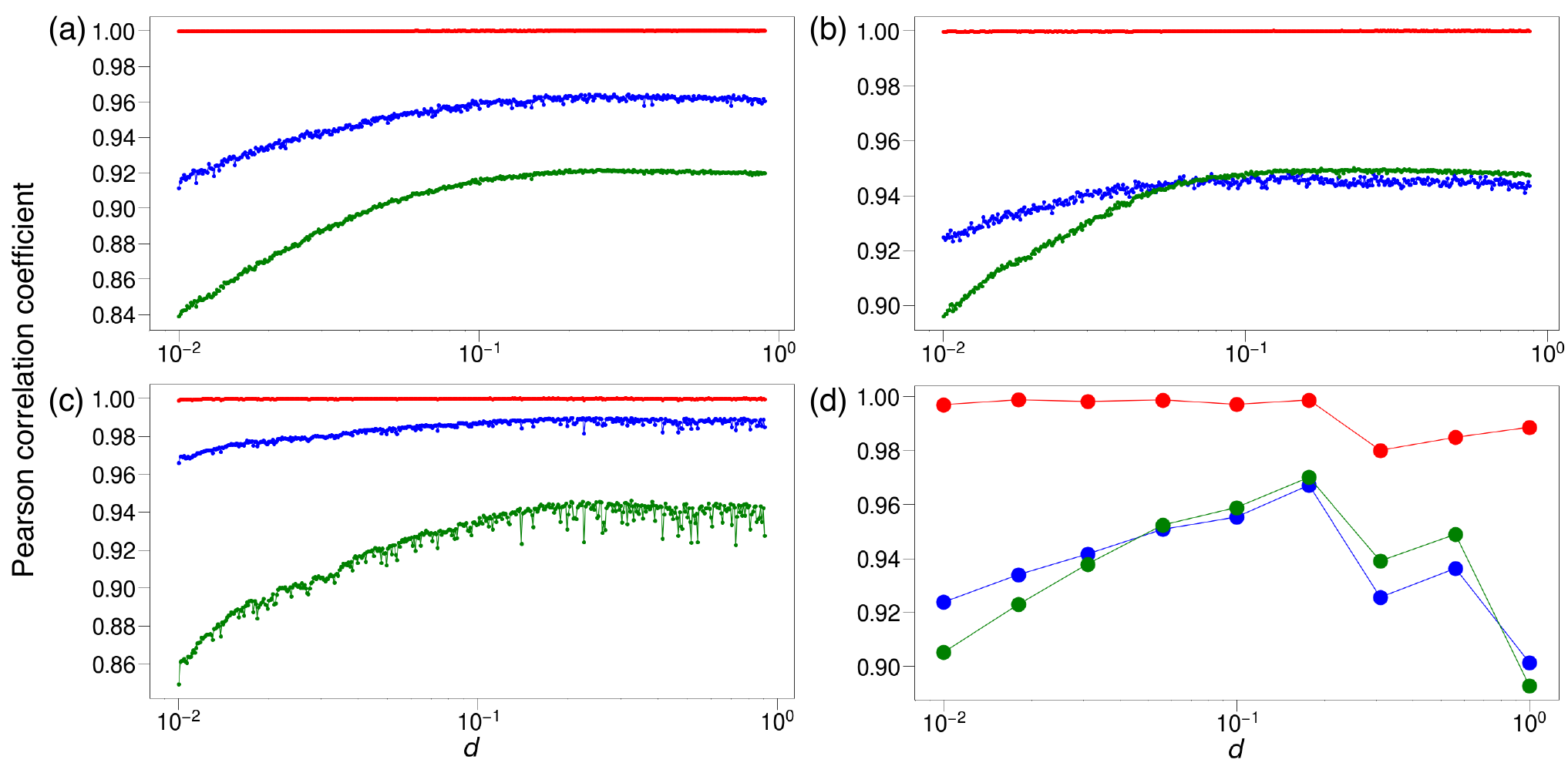}}
\caption{Pearson correlation coefficients between the three chosen metrics (first moments, two-site and three-site connected correlations) of the data and the sparse models as a function of the density. Each panel (a), (b), (c) and (d) is associated with a different family, respectively, PF00014, PF00072, PF00595, PF13354.}
\label{fig:corr_all}
\end{figure*}

To evaluate the quality of the sparse models we compute, for each possible density, the Pearson correlation coefficients between a certain type of statistics computed from the empirical data (the MSA) and the model (via MCMC). More precisely, we focus on one-site frequencies and two- and three-site connected correlations, as defined in Eq.~\eqref{eq:c3}.
To select the indices of the most significant three-site connected correlations we have first extensively scanned all possible triplets  and computed the empirical frequencies for all possible color assignment. We then keep the elements $c_{ijk}(a,b,c)$ with empirical absolute values above $10^{-4}$: only for those elements we compute the corresponding model correlations, in order to limit the computational cost. 
The model correlations are computed from a set of samples generated via the MCMC procedure described in Appendix~\ref{sec:MCMC}. 

In Fig.~\ref{fig:corr_all}, we show the Pearson correlation coefficients for the three metrics between the data and the models as a function of the model density, for the PF00014 (a), PF00072 (b), PF00595 (c) and PF13354 (d) families. For all families, the Pearson coefficient maintains almost the same value reached for the densest (fully connected) model up to a density of about 10\%. When the density goes below 10\%, the Pearson coefficient gradually decreases for all families; not surprisingly, the reduction associated with the three-site connected correlations is more pronounced, because this more-than-two-site correlation is not explicitly fitted by BM learning.

The case of PF13354 is special because, for the reasons discussed in Appendix~\ref{sec:MCMC}, the learning, which is done using rather short waiting times between samples, suffers from a very long decorrelation time in the dense case. Hence, the resampling degrades when MC chains are evolved for long times, which explains why the Pearson coefficients are poor for $d>20\%$. For
$d<20\%$, the decorrelation time becomes much shorter, and the resampling is stable over time, but the Pearson coefficients get progressively degraded when $d$ is reduced, as for the other families. The optimal compromise seems to be $d\approx 20\%$ for this atypical family. See Ref.~\cite{decelle2021equilibrium} for a more detailed analysis of these non-equilibrium sampling effects.

\subsection{Contact prediction}

\begin{figure*}[t]
\centering
 \makebox[\textwidth]{\includegraphics[width=1.00\linewidth]{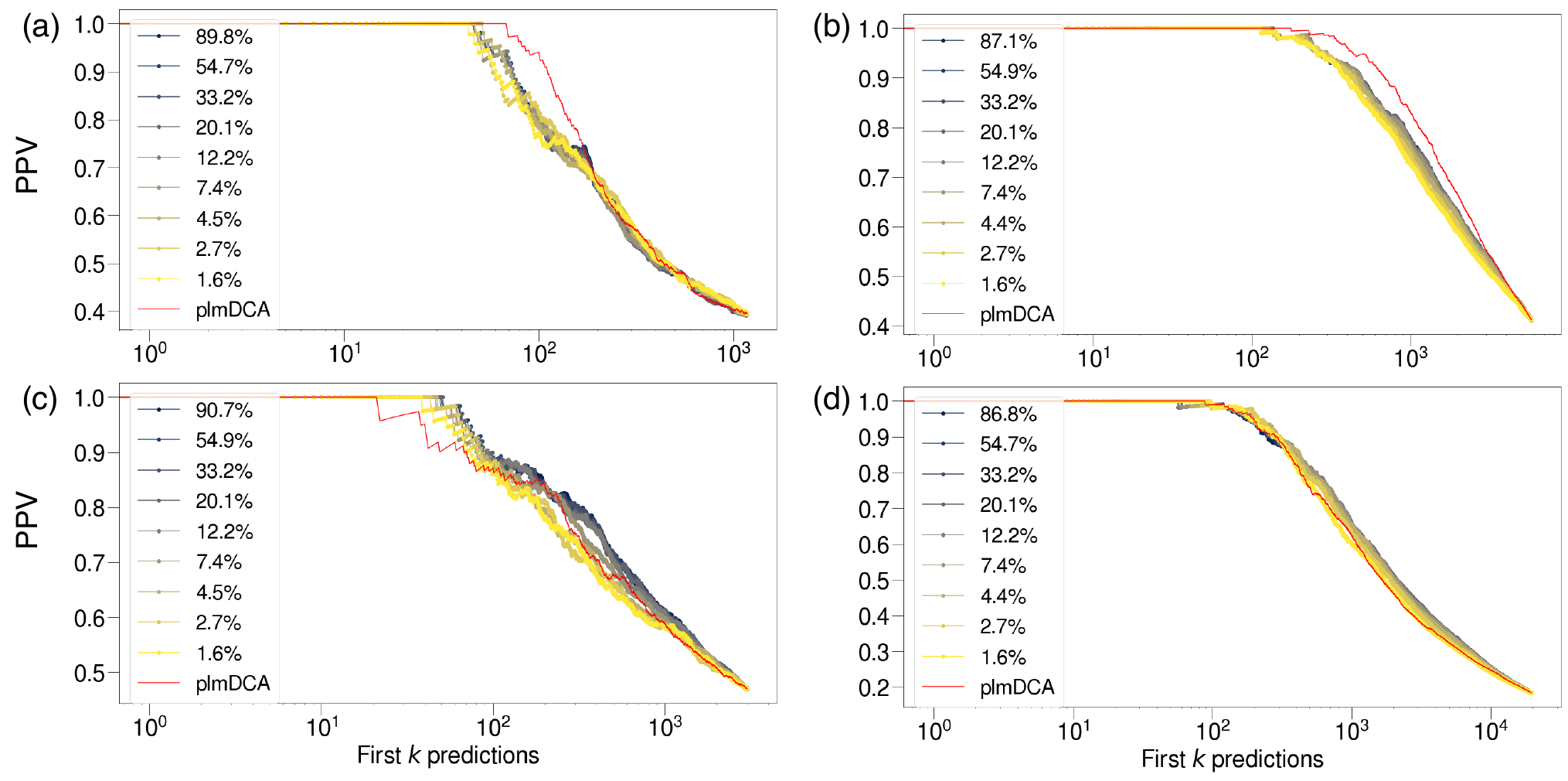}}
\caption{Positive predictive value (PPV) curve for (a) PF00014, (b) PF00072, (c) PF00595 and (d) PF13354 associated with the contact prediction of several sparse models, from yellow to black lines. As a comparison we show the PPV curve (red line) obtained by the state-of-the-art method for this task, \texttt{plmDCA}.
}
\label{fig:ppv_all}
\end{figure*}

The APC-corrected Frobenius norms associated with the couplings can be used for scoring each pair of sites of the MSA (\cf main text). As already explored in literature, this score correlates well with the physical distances between pairs of residues in the three-dimensional structure of the protein domains. Larger Frobenius norms suggest larger probabilities of a physical interaction. As usual, we try to estimate the quality of the sparse models through a set of Positive Predictive Value (PPV) curves associated with the prediction of contacts. As reference structures we use those extracted from \cite{edoardo_sarti_andrea_pagnani_2020}, a tool that outputs the shortest relative distance of pairs of residues over all known crystal structures registered in the Protein Data Bank (PDB) database~\cite{berman2007worldwide}. 
In Fig.~\ref{fig:ppv_all}, we show the PPV curves for a sub-set of the sparse models (the density is mapped to a different color of the lines) together with the result of \texttt{plmDCA}~\cite{ekeberg2014fast} used here as comparison (red lines). 
Even keeping only 10\% of the coupling parameters, i.e. when 90\% of them are removed by the decimation procedure, the accuracy of the contact prediction remains stable, that is the performances are comparable to those of the fully connected models.
The comparison to \texttt{plmDCA} is instead heterogeneous: as found in~\citep{figliuzzi2018pairwise}, the Boltzmann machine learning can have comparable performances to \texttt{plmDCA} as for PF13354 in panel (d), slightly worse as for PF00014 and PF00072 in panels (a), (b), or slightly better as for PF00595 in panel~(c).

\subsection{Coupling distribution}

 Because the couplings mirror a physical interaction among residues, one may guess that the more we decimate the model, the more we decimate the couplings not associated with residues in contact. Similarly, one may expect that the more a coupling is important in terms of three-dimensional structure, the larger will be its strength, hence it will be preserved by the decimation.

\begin{figure*}[t!]
\centering
 \makebox[\textwidth]{\includegraphics[width=1.0\linewidth]{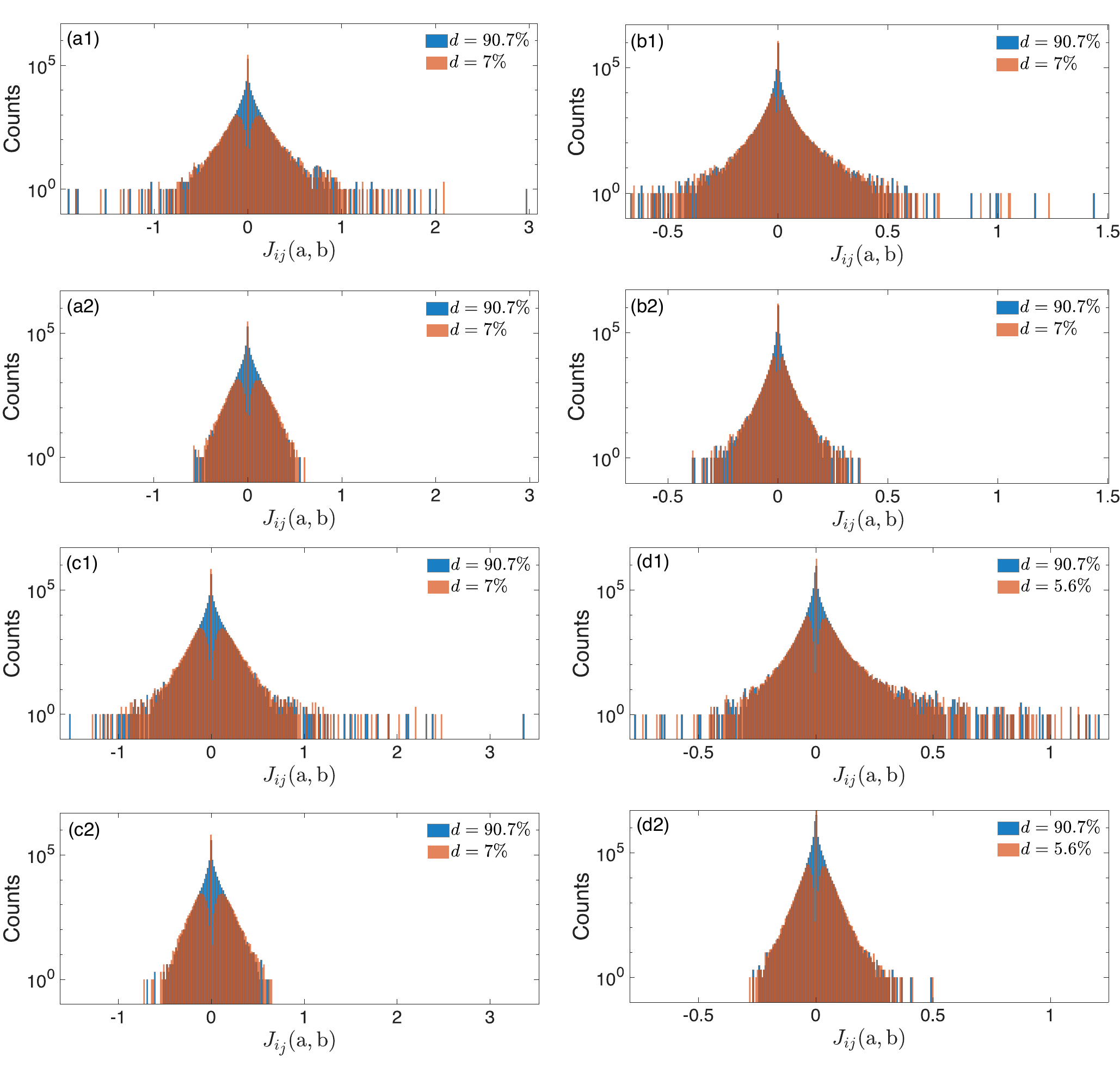}}
\caption{Distribution of the couplings associated with residues physically in contact (labeled `1` histograms) and with residues not in contact (labeled as `2` histograms) for two different densities. Panels (a), (b), (c) and (d) refer to PF00014, PF00072, PF00595 and PF13354 respectively.
\label{fig:JSD_all}
 }
\end{figure*}

To check whether this is the case, we plot in Fig.~\ref{fig:JSD_all}, for PF00014~(a), PF00072~(b), PF00595~(c) and PF13354~(d), the distributions of the couplings linking residues in contact (panel 1) and not in contact (panel 2); the values of the corresponding densities are indicated in the legend. We note that as we reduce the density of the couplings, those corresponding to residues in contact are slightly enhanced (indeed, the original red histograms in Fig.~\ref{fig:JSD_all} for the dense models are shifted to slightly larger values in the sparse case), but we do not observe a significant change in the tails of the distributions, as discussed in the main text.

\subsection{Criticality}
\label{app:criticality}

Dense Potts models are generally very sensitive to a perturbation of their model parameters: a slight change of the couplings or the fields leads to a dramatic transformation of the model statistics, which thus seems to be close to a phase transition, i.e. to be \textit{critical}. A good measure of the criticality of statistical models is represented by the heat capacity,
which is obtained by applying a global variation to the parameters, $J \to J/T,\  h \to h/T$, 
and measuring the derivative of the average internal energy with respect to the temperature, 
\begin{eqnarray}
    C(T) &=& \frac{\partial \langle H \rangle_{T} }{\partial T}
        = \frac{1}{T^2}\Big( 
        \langle H^2 \rangle_{T} - \langle H \rangle^2_{T} 
        \Big) \ .
    \label{eq:heat_capacity}
\end{eqnarray}
The averages in Eq.~\eqref{eq:heat_capacity}, denoted as $\langle . \rangle_T$, are evaluated by sampling a system with Boltzmann weight $\exp\{-H/T\}$. 
Standard thermodynamic identities also show that $T C(T) = 
\partial S/\partial T$, where $S(T)$ is the entropy of the model.
The model criticality
is related to the magnitude of $C(T)$ in the vicinity of $T=1$,
which expresses how quickly the model entropy (or energy) varies under a small rescaling of all couplings. 

\begin{figure*}[t!]
\centering
 \makebox[\textwidth]{\includegraphics[width=1.0\linewidth]{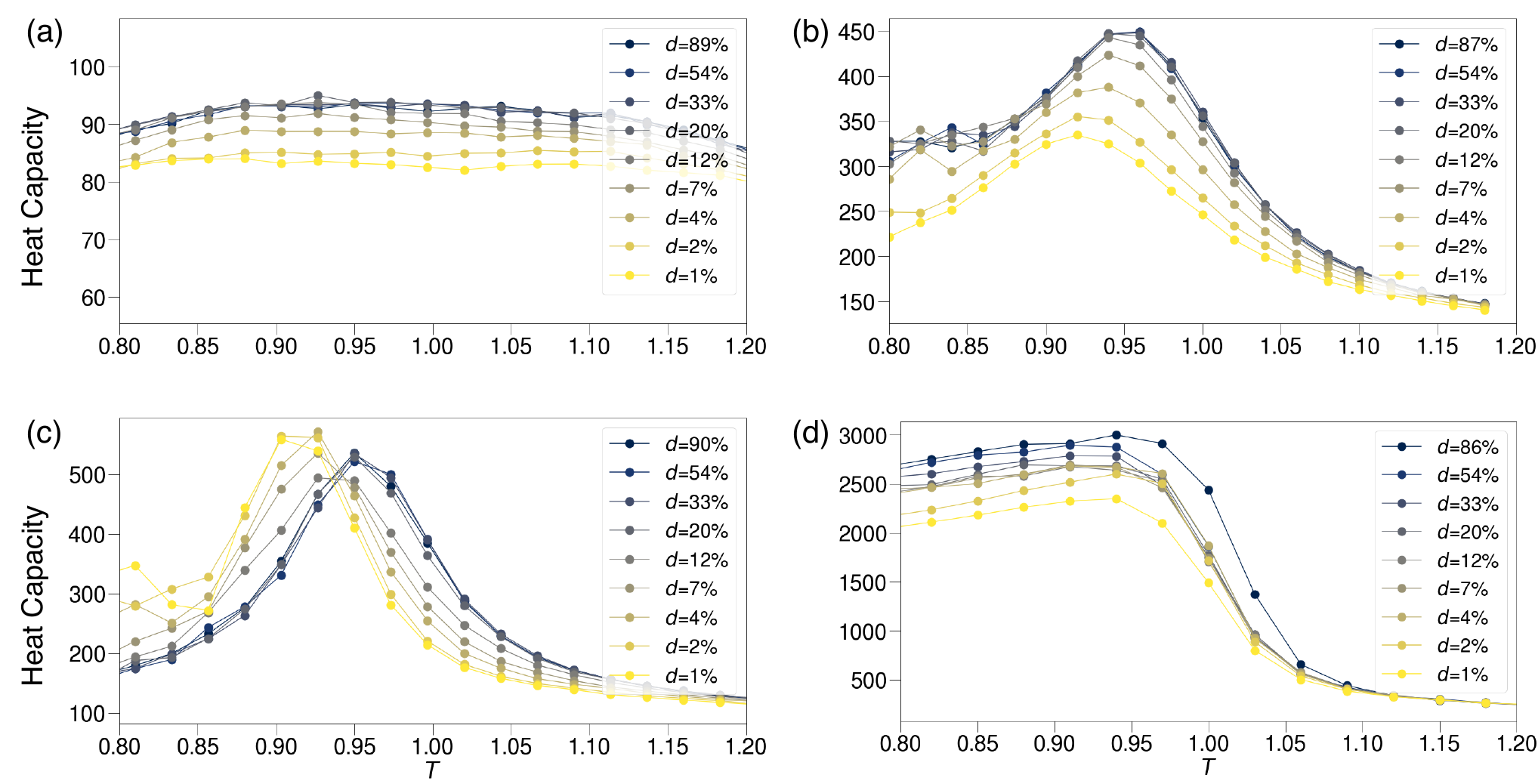}}
\caption{Heat capacity as a function of the temperature $T$ for the other protein families PF00014, PF00072, PF00595 and PF13354 in panels (a), (b), (c) and (d) respectively. 
}
\label{fig:HC_all}
\end{figure*}

Fig.~\ref{fig:HC_all} shows the behavior of the heat capacity $C(T)$ as a function of the temperature $T$ for the models associated with the four families analyzed here: the color of the lines depend on the value of the density of the corresponding model, which spans the range $(1,\,90) \%$.
We observe that for all families, upon sparsifying the model, (i) the heat capacity is reduced rendering the model less sensitive to changes in the model parameters and/or (ii) the peak slightly shifts towards a temperature smaller than $T = 1$, the natural temperature of the learning. In all cases, the value of $C(T=1)$ decreases upon sparsifying the model.
This observation suggests that a dense model learned by the empirical data is indeed close to a phase transition, but the criticality disappears (or decreases substantially) for the statistically equivalent sparser models. Hence, we conclude that the sensitivity of the dense model is related to over-fitting. Note that the suppression of criticality is also suggested by the reduction of the decorrelation time, as discussed in Appendix~\ref{sec:MCMC}.

\begin{figure}[t]
\centering
\includegraphics[width=\columnwidth]{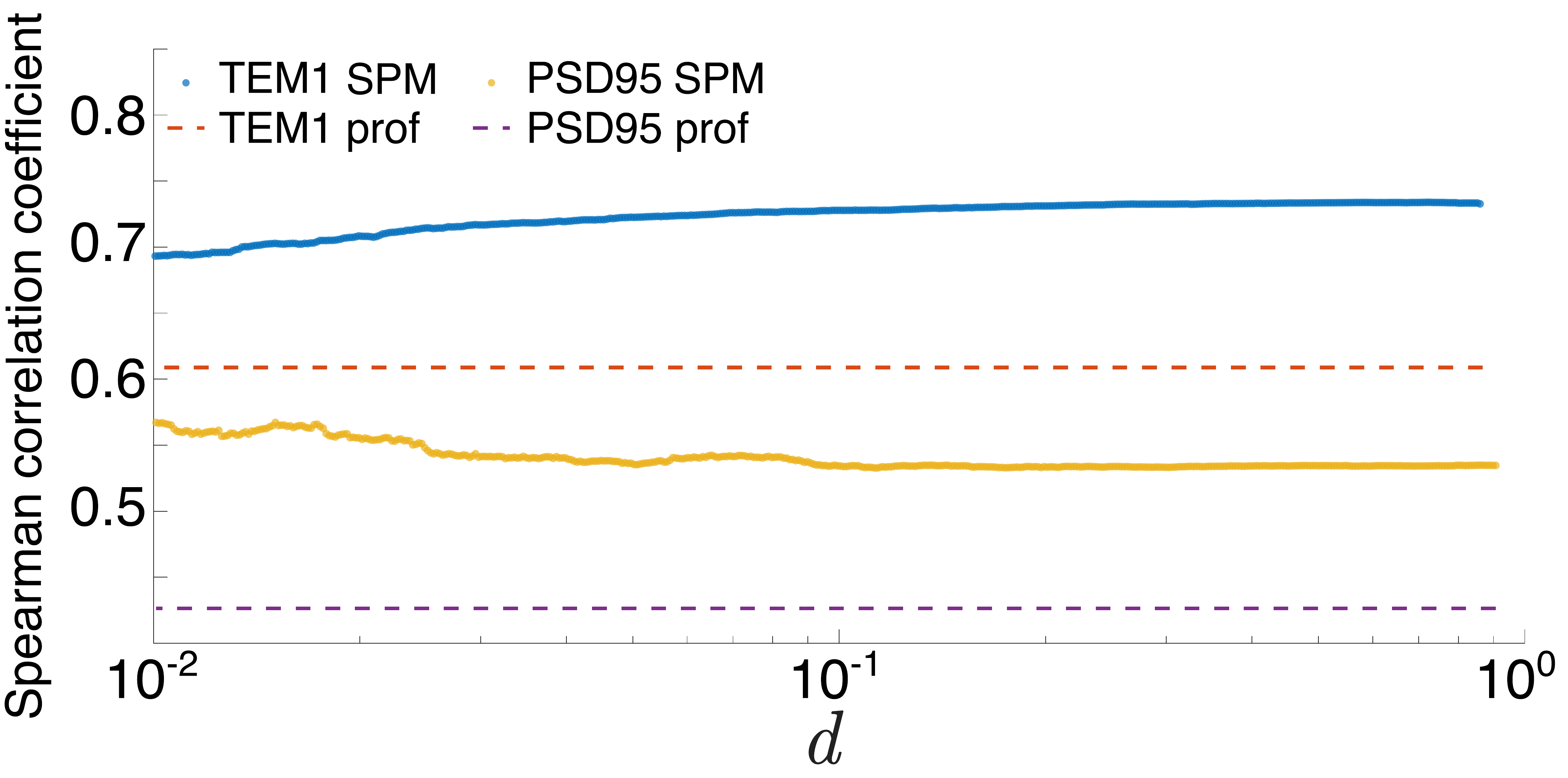}
\caption{Spearman correlation coefficient between the energy variations computed according to the sparse models (spBM lines) and the experimentally determined fitness variations of a set of single mutants, for the TEM1 (PF13354) and PSD95 (PF00595) proteins. The dashed lines show the results of the Spearman correlation coefficients when the energy variations are computed by the profile models of the corresponding families.
\label{fig:mut_SI}}
\end{figure}

\subsection{Mutational landscape prediction}

Similarly to the analysis we proposed in the main text for the PF00076 landscape and the experimentally determined single and double-mutants fitness, we show in Fig.~\ref{fig:mut_SI} the Spearman correlation coefficient, as a function of the density, between the energy variation (computed according to our models) and the experimental fitness associated with single-residue mutations. Here we consider the libraries of single mutants for the Beta-lactamase2 domain of the TEM1 protein~\cite{firnberg2014comprehensive} 
(here the fitness is related to antibiotic resistance)
and for the PDZ3 domain of the 
PSD95 protein~\cite{mclaughlin2012spatial} (here the fitness refers to the \textit{CRIPT} ligand), which we assume to be described by the models for PF13354 and PF00595 families, respectively.
As shown in Fig. \ref{fig:mut_SI}, the correlation coefficient ({\it spBM} lines) between the experimental measures and the energy differences of our models are mostly constant as a function of the density; only a smooth increment (drop) is appreciated for densities smaller than $10^{-1}$ for PSD95 (TEM1). We remark that even in the sparsest case, the Spearman correlation coefficient never crosses that obtained from a pure profile model (denoted as \textit{prof}) suggesting that the remaining non-zero couplings of our sparse models are fundamental for the good description of the fitness landscape.

\section{Additional results on PF00076}

To complete the analysis described in the main text, we propose here a set of additional results for the PF00076 family. More precisely, we compare the learning and decimation strategy used in the main text and in Appendix~\ref{app:otherproteins} (initialize the parameters in the profile model, learn a dense model until convergence, then perform decimation) to several different initializations of the learning and to other decimation strategies based on different metrics. We also investigate the nature of the decimated couplings, via the statistics of the second moments associated with them, to stress the non-trivial nature of the symmetric Kullback-Leibler based decimation.

\subsection{Decimation strategies}
\label{app:decimationstrategies}
 
The method presented in the main text uses as criterion (or score) for the iterative decimation an information-theory based measure, the symmetric Kullback-Leibler divergence (symKLD) between the model with or without a certain coupling. As a result, the decimation score of each coupling takes into account both its statistical relevance (related to the second moments associated with it), and the strength of the coupling alone. We compare here the results presented in the main text to two simpler strategies where at each decimation step a) we remove 1\% of the weakest couplings or b) we remove 1\% of the couplings associated with the lowest, hence less statistically significant, two-site frequencies.

\begin{figure*}[t]
\centering
\includegraphics[width=1.0\textwidth]{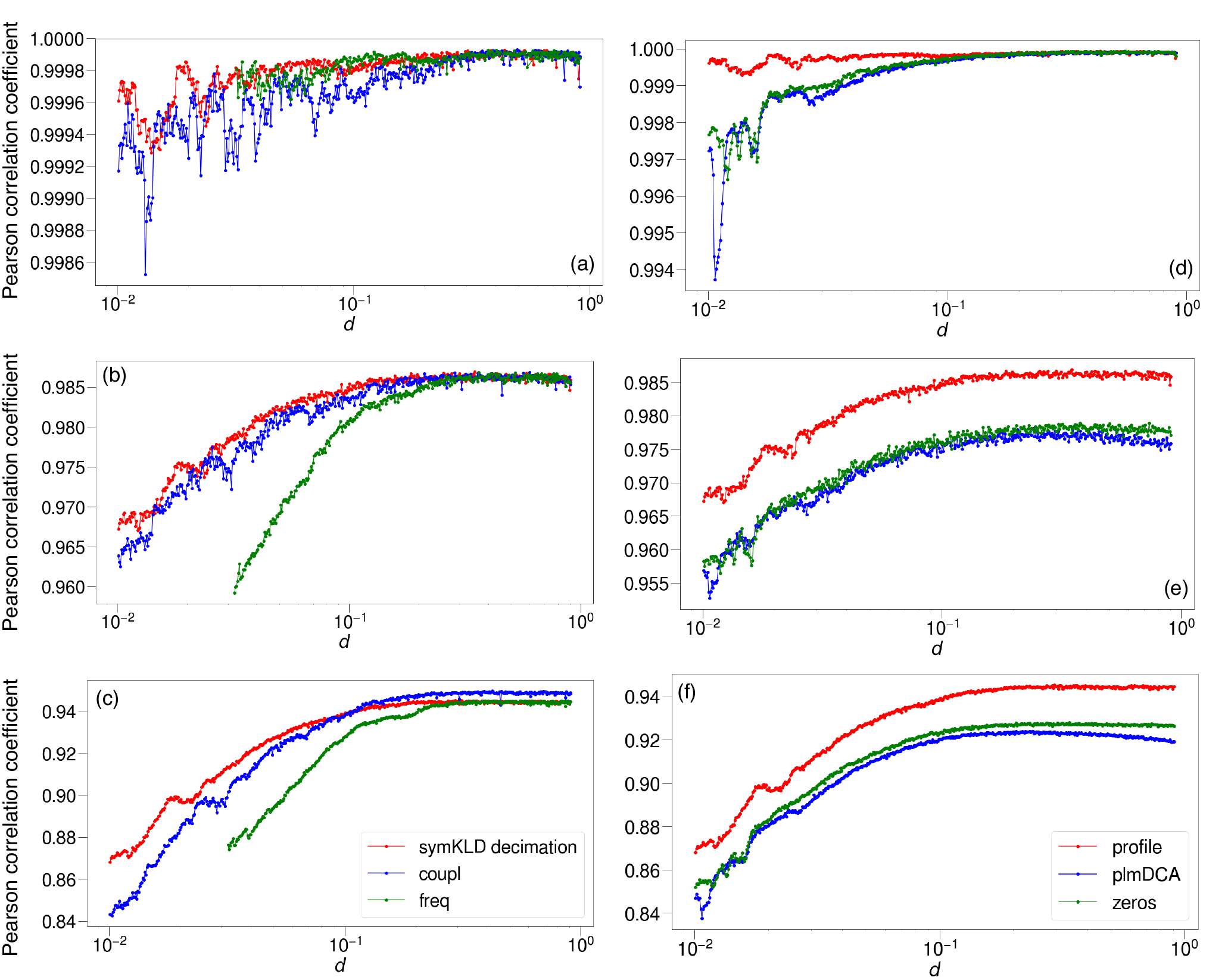}
\caption{Left panels: Pearson correlation coefficients between (a) the first moments, (b) the two- and (c) three-site connected correlations of each model (varying the density) compared to the empirical data. The lines are colored according to the metrics used within the decimation procedure: the symmetric Kullback-Leibler distance, the strength of the couplings or the two-site frequencies. Right panels: Pearson correlation coefficients between (d) the first moments, (e) the two- and (f) three-site connected correlations varying the initial condition of the dense model learning.
All data are for the PF00076 family.
}
\label{fig:corr_decimation_procedure}
\label{fig:corr_init_condition}
\end{figure*}

In Fig.~\ref{fig:corr_decimation_procedure}, in the left panels, we 
compare the three possible decimation procedures using as comparison metric the fitting quality of the sparse models. We show the Pearson correlation coefficient of the empirical data and our sparse models predictions, as a function of the density, for the first moments (panel a), the two-site (panel b) and the most relevant three-site (panel c) connected correlations, respectively. 
Among the three procedures, that based on the two-site frequencies gives the poorest results, as it always provides the lowest Pearson up to density $\approx$3\% where the algorithm fails to converge, meaning that it is no more able to fit the statistics associated with the non-zero parameters. The decimation based on the coupling strength outperforms the frequencies-based one but the Pearson coefficients, for all comparison metrics, is systematically lower than that of the symKLD-based decimation.

\begin{figure}[t]
\centering
\includegraphics[width=\columnwidth]{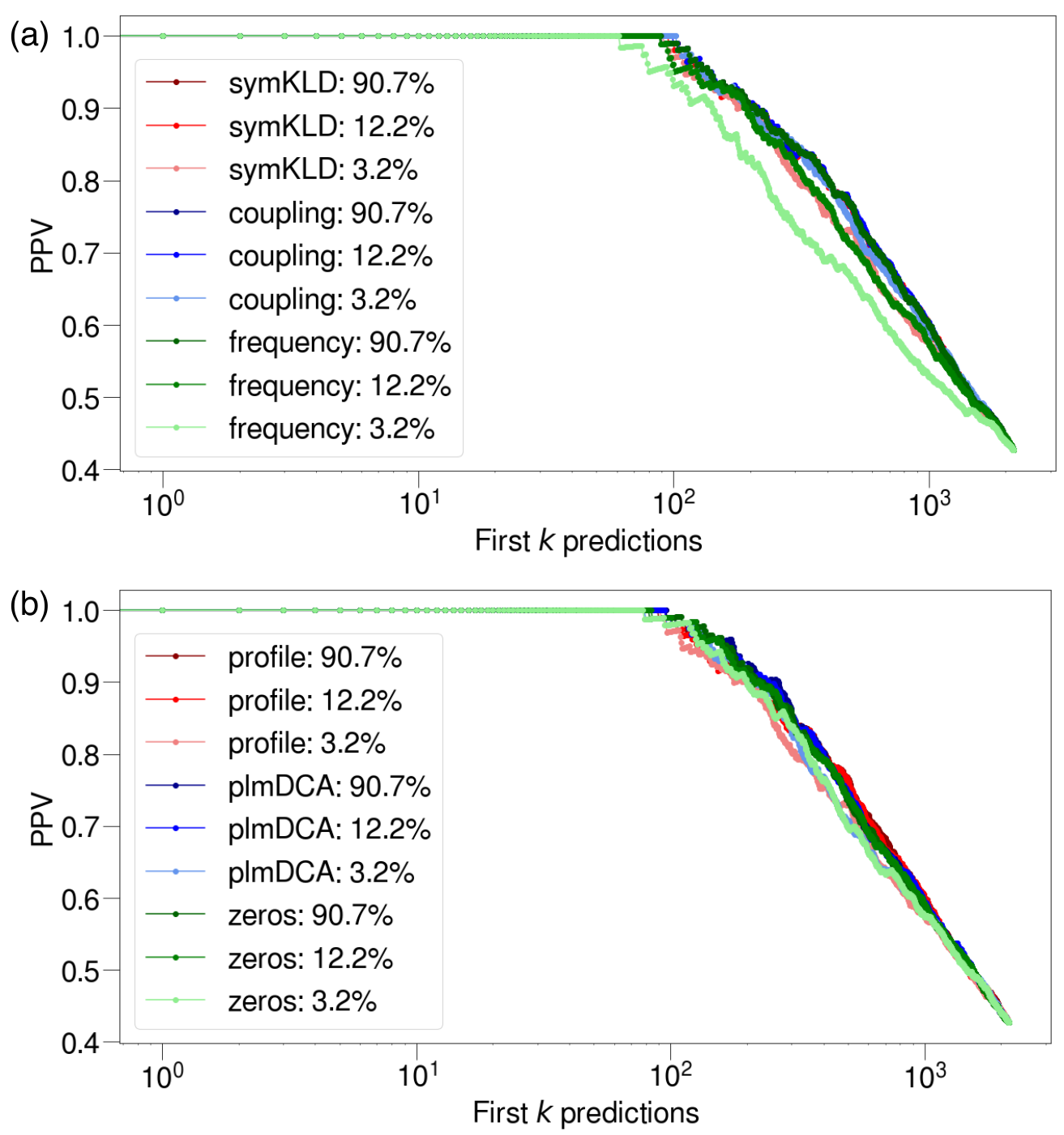}
\caption{Positive predictive value for each decimation procedure (a) and for each initial condition (b) for the PF00076 family.
}
\label{fig:ppv_decimation_procedure}
\label{fig:ppv_init_condition}
\end{figure}

In addition to the fitting quality, we compare the three methods looking at the contact prediction PPV curves, shown in the top panel of Fig.~\ref{fig:ppv_decimation_procedure}, varying the model density. It is worth noting that all procedures, for all densities (except 3.2\% using a frequency-based measure) perform equally well. 

\begin{figure}[t]
\centering
\includegraphics[width=\columnwidth]{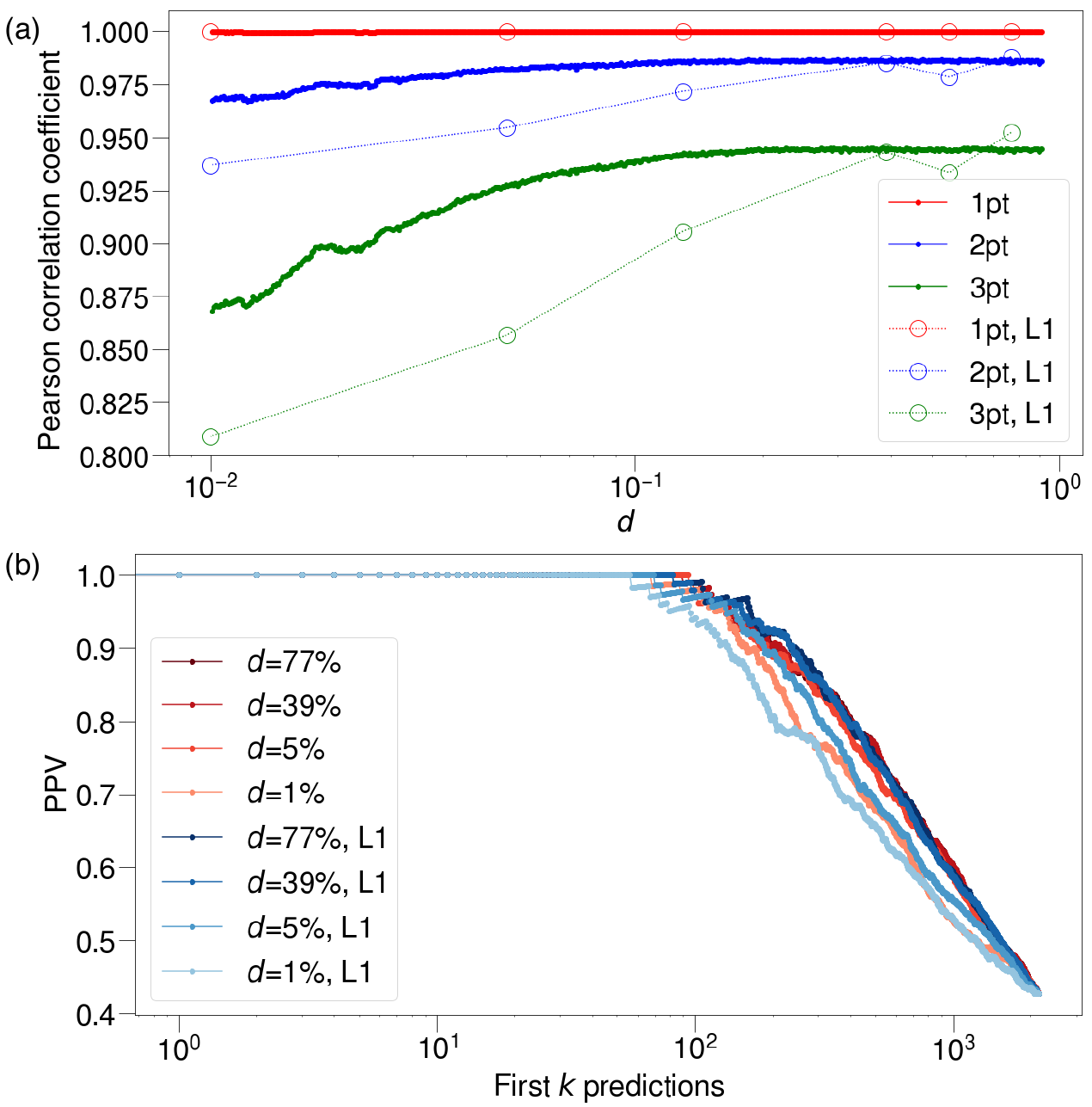}

\caption{
Pearson correlations (a) and
positive predictive value (b) for the decimation via $\ell_1$-norm regularization, for the PF00076 family, compared with those reported in the main text.
}
\label{fig:L1}
\end{figure}

We also considered a standard network selection strategy, in which we first learn a series of dense models with a $\ell_1$-norm regularization at different strength $\gamma$, i.e. Eq.~\eqref{eq:PMlearning} for the couplings is modified to
\beq
\delta J_{ij}(a,b) = \eta_J [ f_{ij}(a,b) - p_{ij}(a,b) ] - \gamma\text{sgn}(J_{ij}(a,b)) \ .
\eeq
At convergence, all couplings such that $|f_{ij}(a,b) - p_{ij}(a,b)|<\gamma$ thus have zero gradient and are considered as decimated. In this way one can obtain PMs of different density $d$ by tuning $\gamma$. After selection, the sparse PMs is trained again keeping the decimated couplings to zero, but without the $\ell_1$-norm regularization for the non-decimated couplings, until convergence. The results for this procedure are reported in Fig.~\ref{fig:L1}, and are outperformed by the symDKL-based procedure.

\begin{figure}[t]
\centering
\includegraphics[width=\columnwidth]{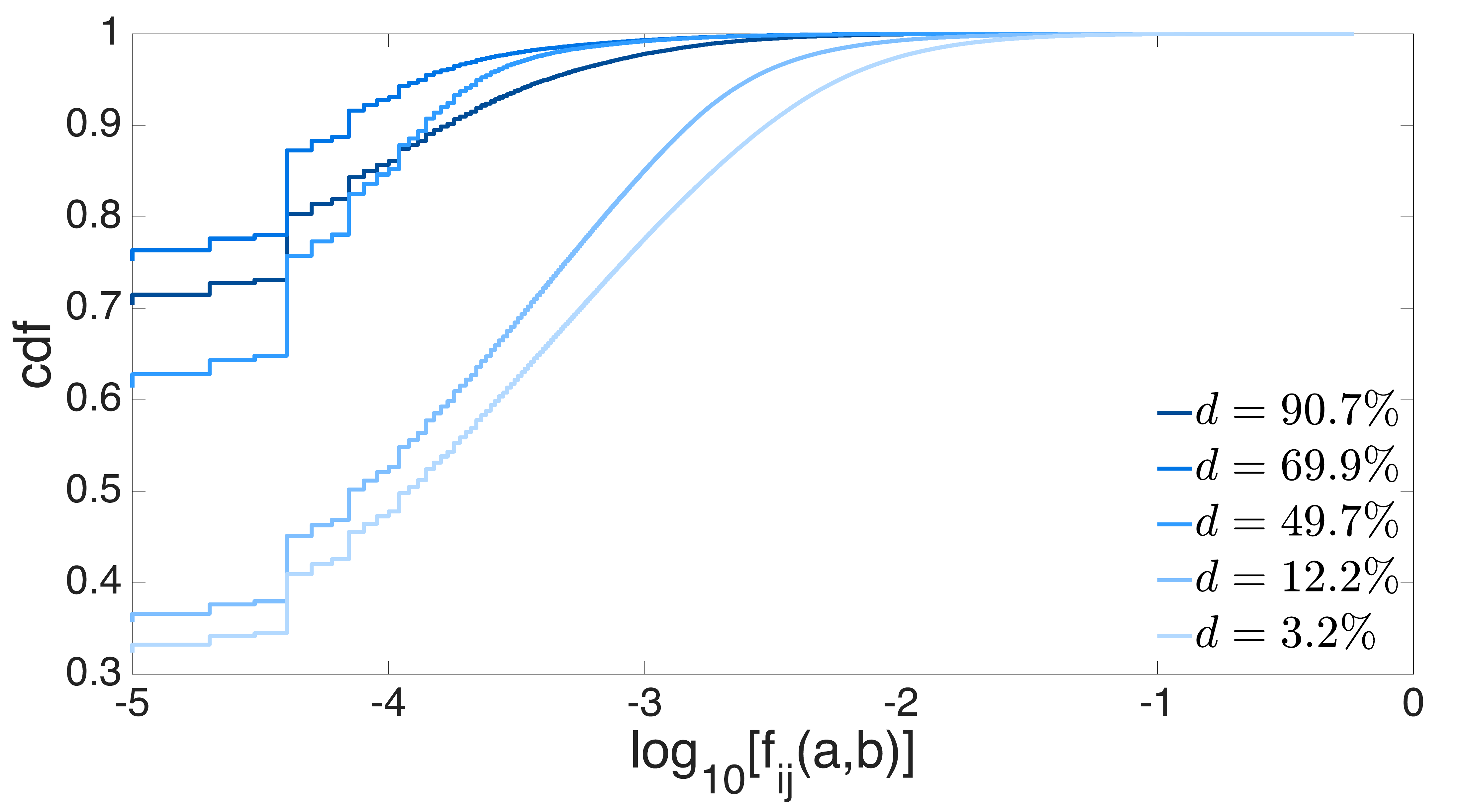}
\caption{Cumulative density function of the logarithms of the two-site frequencies associated with the decimated couplings for the sparse models having densities $d = 90.7 \%$, $d = 69.9 \%$, $d = 49.7 \%$, $d = 12.2 \%$ and $d = 3.2 \%$. The data refers to the PF00076 and the decimation is performed according to the standard protocol described in the main text. \label{fig:cum_fij}}
\end{figure}

\subsection{Decimated couplings}
\label{app:decimatedcouplings}

As mentioned in the previous section, the couplings that are decimated at each iteration are either associated with poor statistics, i.e. pairs of residues that are rarely or very frequently observed in two specific positions, or their strength is very small rendering their contribution in the Boltzmann weight negligible. It is interesting to quantify how many decimated couplings fall into the first or second class, as a function of the density. To this purpose we plot in Fig.~\ref{fig:cum_fij} the empirical cumulative density function of the (logarithms of) the two-site empirical frequencies associated with the decimated couplings. We report in the same plot several curves depending on the density of the considered model: more specifically we observe the cases $d \in \{90.7, 69.9, 49.7, 12.2, 3.2\} \%$.  The values of $\log_{10}[f_{ij}(a,b)]$ in the range $[-5, -4.3]$ empirically correspond to pairs of residues $(a,b)$ appearing one time in position $(i,j)$. Note that, although these frequencies are associated with a single occurrence, they span an interval, i.e. they are not always equal to the same value, because their computation takes into account the re-weighting protocol described in Ref.~\citep{morcos2011direct}, in which each sequence may have a statistical weight smaller than one.
Therefore, the value of the cumulative density function in $\log_{10}[f_{ij}(a,b)] = -5$ gives the fraction of decimated couplings associated with the pairs $(a,b)$ that are never observed in sites $(i,j)$. We see that this quantity changes as a function of the density: when the model is quite dense (for values of $d = \{90.7, 69.9, 49.7\} \%$) about 70 \% of the decimated couplings corresponds to never observed statistics and thus only 30 \% are associated with negligible couplings. As the model becomes sparser and sparser the fraction reduces and reaches about 30 \% for the sparsest models: here about 70 \% of the decimated couplings are associated with a rich statistics but nonetheless their contribution to the Boltzmann weight is negligible.

\subsection{Initialization of the learning for the dense Boltzmann machine}
\label{app:init}

An intrinsic difficulty arises when comparing statistical models for protein sequences: the set of parameters that are able to reproduce the empirical statistics well and also give a good contact prediction is not unique. Therefore, giving a clear interpretation of the fields and the couplings of the inferred Potts model, i.e. to detect which variables are sufficient to characterize the target ensemble of protein sequences, is a challenging task.  
When the sufficient set of observables is not known, and one attempts to fit all possible pairwise couplings and single-site statistics through the Boltzmann machine learning, it is common to encounter `flat' directions of the log-likelihood landscape, where the learning usually converges (as any attempt at modifying the parameters does not lead to any significant improvement). The parameters found at convergence thus strongly depend on the initial conditions.

Here we evaluate how the results of the decimation procedure are affected by the dense model used as starting point, which in turn depends on the initial conditions of the parameters. For this comparison, we consider three distinct initial conditions for the initial learning of the dense model: (a) the profile model ($h=h^{\rm profile}, J=0$,  used for the results presented in the main text), (b) the parameters from pseudo-likelihood maximization ($h=h^{\rm plmDCA}, J=J^{\rm plmDCA}$), as implemented in \texttt{plmDCA}  \citep{ekeberg2014fast}, and (c) a null initial condition for all model parameters ($h=0, J=0$). We then let the Boltzmann machine learning converge, and we use the converged Potts model as the starting model of the decimation run described in the main text. 

In the right panels of Fig.~\ref{fig:corr_init_condition} we show the Pearson correlation coefficients between the empirical frequencies $f_{i}(a)$ and the model frequencies $p_{i}(a)$ (in panel (a)), and the two-site and three-site connected correlations of the empirical data and of the sparse models, for panels (b) and (c) respectively. When all parameters are initialized to {\it profile} we reach the larger Pearson correlation coefficients, for all the three measures and for all densities. The {\it plmDCA} and \textit{zeros} initializations have comparable results, and they reach Pearson correlation coefficients equal to those of the \textit{profile} initialization only for the first moment in the high density regime.

In addition to the fitting quality, we can compare the three different initializations through the contact map prediction. We observe in Fig.~\ref{fig:ppv_init_condition} that all the three strategies, independently of the density, provide very similar contact prediction as the associated PPV curves completely overlap.

\begin{figure}[t]
    \centering
    \includegraphics[width=0.5\textwidth]{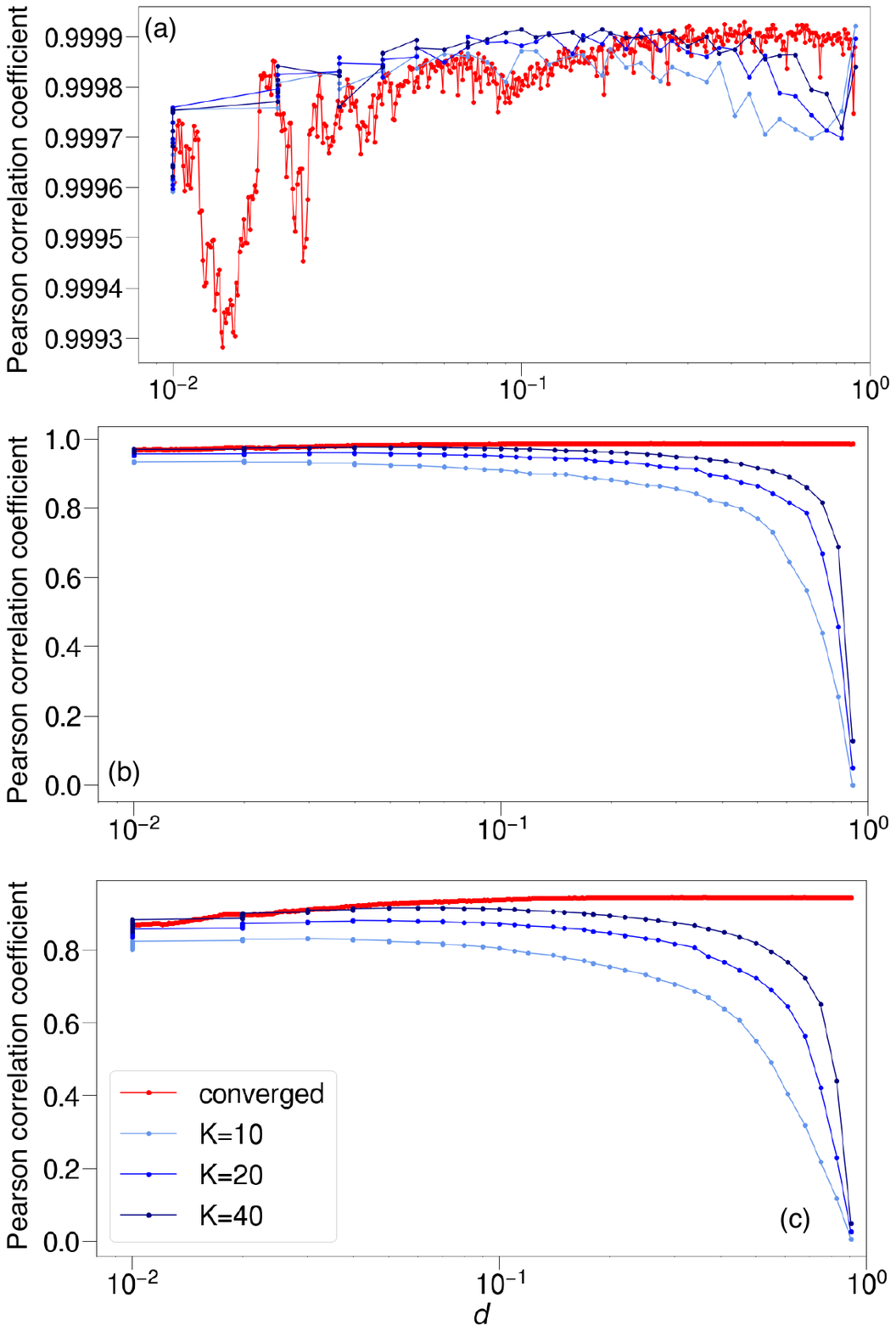}
\caption{Pearson correlation coefficients for the \textit{online} learning (blue line), for $K = 10, 20, 40$, compared to the \textit{converged} run (red line) as a function of the density, for the one-site frequencies (a), two-site (b) and three-site (c) connected correlations.}         \label{fig:corr_online}
\end{figure}

\begin{figure}[t]
    \centering
\includegraphics[width=0.5\textwidth]{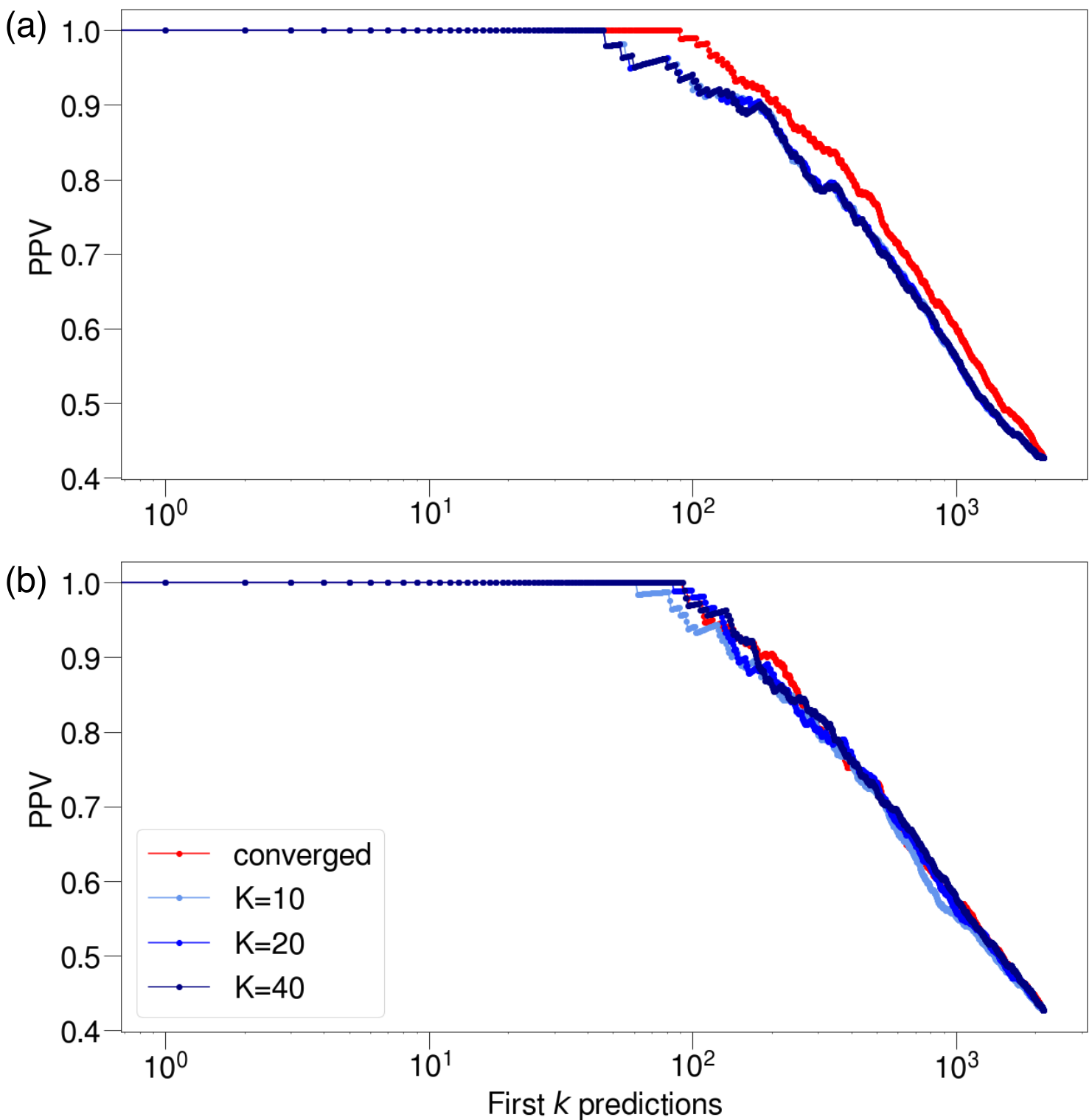}
    \caption{Comparison between the Positive Predictive Value (PPV) curve obtained for the \textit{online} (for $K = 10, 20, 40$) and \textit{converged} runs at two different densities, 72\% (a) and 3.2\% (b). All data are for the PF00076 family. }
    \label{fig:ppv_online}
\end{figure}

\subsection{Online learning}
\label{app:onlinelearning}

In our decimation protocol, we proceed with a new decimation step only when the learning has reached convergence. Starting from a well converged dense Potts model, and decimating only 1\% of the couplings at the time, allows us to modify smoothly the remaining parameters during the decimation. Indeed, we empirically observe that most of the times two consecutive decimations are separated by just a few learning steps.
However, the entire protocol requires to learn a dense model first, which can be time-consuming.

We thus explored an alternative strategy in which the decimation is performed on-line, i.e. within a unique learning run. Here the decimation step is applied either because the learning has performed $K$ steps or because it has reached the tolerance required for convergence. 
In these experiments, we start from a set of parameters corresponding to the profile model (as in the protocol illustrated in the main text) for PF00076 and we proceed with the decimation step every $K = {10, 20, 40}$ steps. 

In Fig.~\ref{fig:corr_online} we compare the Pearson correlation coefficient between the one-site frequencies (panel a), two-site (b) and three-site (c) connected correlations, of the data and the models obtained by the two different strategies: we refer to the conventional method as \textit{converged} (corresponding to $K\to\infty$) while the on-line learning method is characterized by the number $K$ of steps. It is worth noting that, at convergence, both strategies, and independently of $K$, reach the same fitting quality even in the three-site connected correlations.
For completeness, we show in Fig.~\ref{fig:ppv_online} the contact prediction performance of the \textit{converged} and on-line runs for densities equal to 72\% (panel a) and 3.2\% (panel b). In the denser case (when we consider 72\% of non-zero couplings) the converged run outperforms the on-line learning for any $K$. This can be explained by the poor fitting quality reached by the on-line runs at the initial steps of the algorithm, that is when the model is still inaccurate in fitting the two-site frequencies. It is worth noting that, in the sparse regime, i.e. for density equals to 3.2\%,  all the strategies show comparable results, qualitatively similar to the performance of the dense case (Fig.~\ref{fig:ppv_online}a).  

Although the results of the on-line run resemble those of the converged run for the very sparse models, the on-line procedure is not always advantageous from the point of view of the running time. We notice that, depending on the family, a unique learning-decimation run may have problems fitting the statistics, i.e. to converge, because the decimation affects and `deviates' the learning of the machine, for small $K$. To cure this issue, one may think of increasing the number of steps between each decimation. However, this results in a very slow procedure, because we remove 1\% of the couplings every (large) $K$ steps. Instead, if the model is well converged first, then convergence is achieved quite fast after each decimation, resulting in a faster procedure overall.

\section{Sequences similarity}
\label{app:similarity}

\begin{figure*}[t]
\centering
 \makebox[\textwidth]{\includegraphics[width=1.0\linewidth]{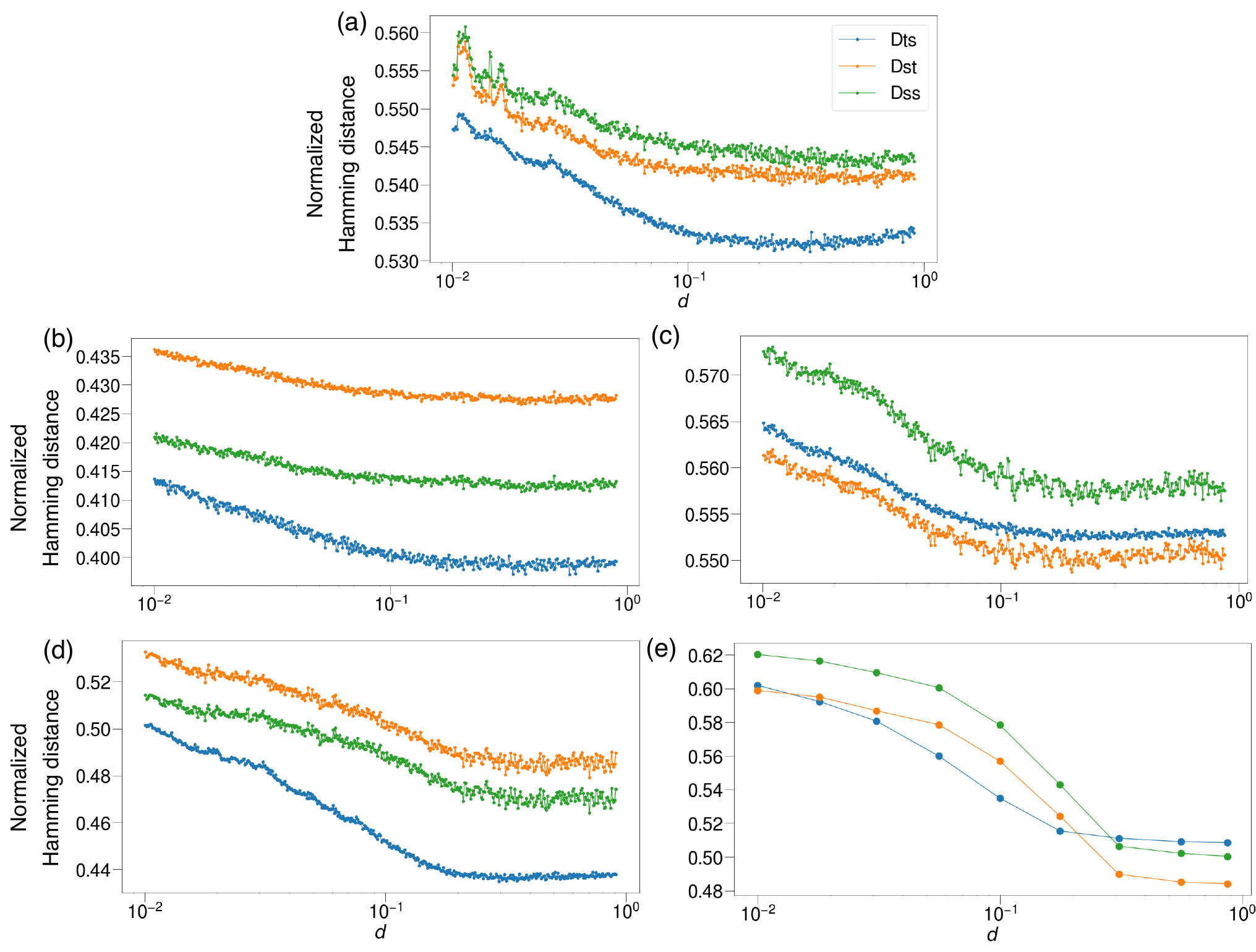}}
\caption{Sequence variability for all the protein families considered in this work. We plot $D_{\rm ts}$, $D_{\rm st}$ and $D_{\rm ss}$ as a function of the density using blue, orange and green lines, respectively, for PF00076 (panel a), PF00014 (b), PF00072 (c), PF00595 (d), PF13354 (e).
}
\label{fig:Sequence_variability}
\end{figure*}

The defining feature of generative models is the ability to generate configurations that are statistically equivalent to those used within the training process, but substantially different in the residue composition, i.e. a good generative model should not just reproduce the sequences of the training set.
Hence, it is important to quantify the distances between generated samples and the training data.
For this purpose, we employed the following metrics, introduced in \citep{yale2019privacy, yelmen2019creating}:
\begin{equation}
D_Y(x) = \min_{y\in Y}\ D(x, y) \ , \qquad
D_{XY} = \frac{1}{N_X}\sum_{n=1}^{N_X} D_Y(x_n) \ .
\end{equation}
where $X$ and $Y$ are ensembles of the generic statistical variables $x$ and $y$, $D(x, y)$ is a certain distance defined for the sequences $x$ and $y$. The metric $D_Y(x)$ computes the minimum distance of the sequence $x$ reached when compared to each of the possible sequences in the ensemble $Y$; the quantity $D_{XY}$ is instead the average value of $D_Y(x)$ over the ensemble of $X$. In our problem, we choose $D(x,y)$ as the Hamming distance between sequence $x$ and sequence $y$ and the ensembles $X$ and $Y$ are respectively $t$ (the training set) and $s$, the synthetic sequences generated from the sparse Boltzmann machines. A proper generative model would produce comparable $D_{\rm st}$ and $D_{\rm ss}$ and, concurrently, the two measures must be sufficiently large (practically 20\% of sequence similarity is required for good training sets). This corresponds to a scenario where generated sequences are variable (large $D_{\rm ss}$), and similarly distant to natural or the other generated sequences ($D_{\rm ss}\simeq D_{\rm st}$). This corresponds to a scenario where the average distance between each pair of generated sequences is comparable to that obtained between the two ensembles $t$ and $s$: therefore, the generated synthetic sequences are indistinguishable from the natural sequences using distance based methods (like nearest-neighbor classification, distance based clustering). A similar argument can be applied to $D_{\rm ts}$.
In Fig.~\ref{fig:Sequence_variability}, we show the average distances $D_{\rm ts}$, $D_{\rm st}$ and $D_{\rm ss}$ for each protein family; we do not show the $D_{\rm tt}$ measure which is obviously constant for all densities, and takes values $D_{\rm tt}(\text{PF00076})=0.308$, $D_{\rm tt}(\text{PF00014})=0.0917$, $D_{\rm tt}(\text{PF00072})=0.421$, $D_{\rm tt}(\text{PF00595})=0.295$, and $D_{\rm tt}(\text{PF00076})=0.445$. Because of the phylogenetic relationship among sequences, the training set is composed of similar (correlated) sequences and, as a consequence, the $D_{\rm  tt}$ is significantly smaller than the other distance metrics.
Regarding $D_{\rm ts}$, $D_{\rm st}$ and $D_{ss}$ we notice that, as the density of the couplings decreases, the distances remain unchanged up to a density in the range 10\% - 20\%, depending on the family. Then the minimum average distance significantly increases which suggests that the synthetic sequences are distributed more broadly in the sequence space as the number of model parameters decreases. Besides, the difference between $D_{\rm ts}$, $D_{\rm st}$, and $D_{\rm ss}$ decreases for most of the protein families, in the sparse regime, suggesting that the synthetic sequence ensembles and the set of the natural sequences become more and more statistically similar for increasing sparsity.
We can conclude that, according to these metrics, the decimation improves the generative properties of the model.

\bibliography{sparseDCA.bib}

\end{document}